\documentclass[aps,prd,twocolumn,groupedaddress,showpacs,floatfix,letterpaper,amsmath]{revtex4}

\usepackage{graphicx}

\begin{document}

\def\al{\alpha}
\def\be{\beta}
\def\ga{\gamma}
\def\de{\delta}
\def\ep{\epsilon}
\def\ve{\varepsilon}
\def\ze{\zeta}
\def\et{\eta}
\def\th{\theta}
\def\vt{\vartheta}
\def\io{\iota}
\def\ka{\kappa}
\def\la{\lambda}
\def\vpi{\varpi}
\def\rh{\rho}
\def\vr{\varrho}
\def\si{\sigma}
\def\vs{\varsigma}
\def\ta{\tau}
\def\up{\upsilon}
\def\ph{\phi}
\def\vp{\varphi}
\def\ch{\chi}
\def\ps{\psi}
\def\om{\omega}
\def\Ga{\Gamma}
\def\De{\Delta}
\def\Th{\Theta}
\def\La{\Lambda}
\def\Si{\Sigma}
\def\Up{\Upsilon}
\def\Ph{\Phi}
\def\Ps{\Psi}
\def\Om{\Omega}

\newcommand{\elp}{e^{+}}
\newcommand{\elm}{e^{-}}
\newcommand{\mup}{\mu^{+}}
\newcommand{\mum}{\mu^{-}}
\newcommand{\piz}{\pi^{0}}
\newcommand{\pio}{\pi^\circ}
\newcommand{\pip}{\pi^{+}}
\newcommand{\pim}{\pi^{-}}
\newcommand{\nue}{\nu_e}
\newcommand{\nueb}{\bar{\nu}_e}
\newcommand{\numu}{\nu_\mu}
\newcommand{\numub}{\bar{\nu}_\mu}
\newcommand{\nutau}{\nu_\tau}
\newcommand{\nutaub}{\bar{\nu}_\tau}

\newcommand{\no}{\nonumber}
\newcommand{\beq}{\begin{eqnarray}}
\newcommand{\eeq}{\end{eqnarray}}

\newcommand{\Elo}{E_\mathrm{lo}}
\def\shc{sin^2\left(\fr{1}{2}\ch\right)}
\def\chc{cos^2\left(\fr{1}{2}\ch\right)}
\def\omeff{\om_{eff}}
\def\qeff{q^2_{eff}}
\def\vk{|\vec{k}|}
\def\vq{|\vec{q}|}
\def\vkq{|\vec{k}-\vec{q}|}
\def\kmin{k_{min}}
\def\kmax{k_{max}}
\def\ucm{\mbox{cm}}
\def\ucmt{\mbox{cm}^2}
\def\ucmc{\mbox{cm}^3}
\def\uGeV{\mbox{GeV}}
\def\uMeV{\mbox{MeV}}
\def\ueV{\mbox{eV}}
\def\uGeVt{\mbox{GeV}^2}
\def\ueVt{\mbox{eV}^2}
\def\uMeVct{\mbox{MeV}/\mbox{c}^2}
\def\uGeVct{\mbox{GeV}/\mbox{c}^2}

\def\fr#1#2{{{#1} \over {#2}}}
\def\frac#1#2{\textstyle{{{#1} \over {#2}}}}
\def\pt#1{\phantom{#1}}
\def\prt{\partial}
\def\vev#1{\langle {#1}\rangle}
\def\ket#1{|{#1}\rangle}
\def\bra#1{\langle{#1}|}
\def\amp#1#2{\langle {#1}|{#2} \rangle}
\def\half{{\textstyle{1\over 2}}}
\def\lsim{\mathrel{\rlap{\lower4pt\hbox{\hskip1pt$\sim$}}
    \raise1pt\hbox{$<$}}}
\def\gsim{\mathrel{\rlap{\lower4pt\hbox{\hskip1pt$\sim$}}
    \raise1pt\hbox{$>$}}}
\def\Re{\hbox{Re}\,}
\def\Im{\hbox{Im}\,}
\def\Arg{\hbox{Arg}\,}
\def\etal {{\it et al.}}
\def\obt{\fr{1}{2}}
\def\obrt{\fr{1}{\sqrt{2}}}
\def\nuance{\textsc{nuance}}
\newcommand{\MA}{M_A}
\newcommand{\MAeff}{M_A^{\mathrm{eff}}}


\def\MBosc1POT{5.58\times 10^{20}}
\def\MBnumuflx{5.16\times 10^{-10}}
\def\MBnumuflxint{2.88\times 10^{11}}

\def\PRLMAcon{1.23}
\def\PRLMAerr{0.20}
\def\PRLKAcon{1.019}
\def\PRLKAerr{0.011}

\def\EWTMAcon{1.23}
\def\EWTMAerr{x.xxx}
\def\EWTKAcon{1.022}
\def\EWTKAerr{x.xxx}

\def\NEWMAcon{1.35}
\def\NEWMAerr{0.17}
\def\NEWKAcon{1.007}
\def\NEWKAerr{0.012}

\def\NUAMAcon{1.03} 
\def\NUAMAerr{x.xxx}
\def\NUAKAcon{1.000}
\def\NUAKAerr{x.xxx}

\def\evki{146070}
\def\pinucabs{11}

\def\xfactor_old{1.17}
\def\xfactor_new{1.05}
\def\CHcon{47.0}

\def\absXScon{9.429}
\def\absXSerr{x.xxx}
\def\absXSerrStat{x.xxx}
\def\absXSerrSyst{x.xxx}
\def\itXSerr{0.60}
\def\stXSerr{0.26}
\def\flXSerr{8.66}
\def\xtXSerr{4.32}
\def\deXSerr{4.60}
\def\exXSerr{2.3}
\def\ttXSerr{10.7}

\def\fCCQEt{39}
\def\fNCELt{16}
\def\fCCPIt{25}
\def\fNCPZt{ 8}
\def\fCCPZt{ 4}
\def\fNCPPt{ 4}
\def\fOTHRt{ 4}
\def\fANTIt{<2}

\def\fCCQEs{77.0}      
\def\fNCELs{<0.1}      
\def\fCCPPs{18.4}      
\def\fCCP0s{2.6}      
\def\fNCPPs{0.4}      
\def\fNCP0s{<0.1}      
\def\fOTHRs{0.2}      
\def\fANTIs{1.4}
\def\fTBCKs{23.0} 

\def\eCCQE{112515}
\def\eNCEL{45}
\def\eCCPP{26866}
\def\eCCP0{3762}
\def\eNCPP{535}
\def\eNCP0{43}
\def\eOTHR{328}
\def\eANTI{1977}
\def\eTBCK{33556} 

\def\vetoCCQEeff{54.8}
\def\timeCCQEeff{54.3}
\def\radiCCQEeff{45.0}
\def\enerCCQEeff{39.7}
\def\lemuCCQEeff{36.0}
\def\tsubCCQEeff{29.1}
\def\distCCQEeff{26.6}

\def\vetoCCQEpur{36.8}
\def\timeCCQEpur{36.8}
\def\radiCCQEpur{37.4}
\def\enerCCQEpur{46.3}
\def\lemuCCQEpur{62.3}
\def\tsubCCQEpur{71.0}
\def\distCCQEpur{77.0}

\title{First Measurement of the Muon Neutrino Charged Current \\
Quasielastic Double Differential Cross Section}

\date{\today}

\author{A.~A. Aguilar-Arevalo$^{13}$, C.~E.~Anderson$^{18}$,
	A.~O.~Bazarko$^{15}$, S.~J.~Brice$^{7}$, B.~C.~Brown$^{7}$,
        L.~Bugel$^{5}$, J.~Cao$^{14}$, L.~Coney$^{5}$,
        J.~M.~Conrad$^{12}$, D.~C.~Cox$^{9}$, A.~Curioni$^{18}$,
        Z.~Djurcic$^{5}$, D.~A.~Finley$^{7}$, B.~T.~Fleming$^{18}$,
        R.~Ford$^{7}$, F.~G.~Garcia$^{7}$,
        G.~T.~Garvey$^{10}$, J.~Grange$^{8}$, C.~Green$^{7,10}$, J.~A.~Green$^{9,10}$,
        T.~L.~Hart$^{4}$, E.~Hawker$^{3,10}$,
        R.~Imlay$^{11}$, R.~A. ~Johnson$^{3}$, G.~Karagiorgi$^{12}$,
        P.~Kasper$^{7}$, T.~Katori$^{9,12}$, T.~Kobilarcik$^{7}$,
        I.~Kourbanis$^{7}$, S.~Koutsoliotas$^{2}$, E.~M.~Laird$^{15}$,
        S.~K.~Linden$^{18}$,J.~M.~Link$^{17}$, Y.~Liu$^{14}$,
        Y.~Liu$^{1}$, W.~C.~Louis$^{10}$,
        K.~B.~M.~Mahn$^{5}$, W.~Marsh$^{7}$, C.~Mauger$^{10}$,
	V.~T.~McGary$^{12}$, G.~McGregor$^{10}$,
        W.~Metcalf$^{11}$, P.~D.~Meyers$^{15}$,
        F.~Mills$^{7}$, G.~B.~Mills$^{10}$,
        J.~Monroe$^{5}$, C.~D.~Moore$^{7}$, J.~Mousseau$^{8}$, R.~H.~Nelson$^{4}$,
	P.~Nienaber$^{16}$, J.~A.~Nowak$^{11}$,
	B.~Osmanov$^{8}$, S.~Ouedraogo$^{11}$, R.~B.~Patterson$^{15}$,
        Z.~Pavlovic$^{10}$, D.~Perevalov$^{1}$, C.~C.~Polly$^7$, E.~Prebys$^{7}$,
        J.~L.~Raaf$^{3}$, H.~Ray$^{8,10}$, B.~P.~Roe$^{14}$,
	A.~D.~Russell$^{7}$, V.~Sandberg$^{10}$, R.~Schirato$^{10}$,
        D.~Schmitz$^{5}$, M.~H.~Shaevitz$^{5}$, F.~C.~Shoemaker$^{15}$\footnote{deceased},
        D.~Smith$^{6}$, M.~Soderberg$^{18}$,
        M.~Sorel$^{5}$\footnote{Present address: IFIC, Universidad de Valencia and CSIC, Valencia 46071, Spain},
        P.~Spentzouris$^{7}$, J.~Spitz$^{18}$, I.~Stancu$^{1}$,
        R.~J.~Stefanski$^{7}$, M.~Sung$^{11}$, H.~A.~Tanaka$^{15}$,
        R.~Tayloe$^{9}$, M.~Tzanov$^{4}$,
        R.~G.~Van~de~Water$^{10}$, 
	M.~O.~Wascko$^{11}$\footnote{Present address: Imperial College; London SW7 2AZ, United Kingdom},
	 D.~H.~White$^{10}$,
        M.~J.~Wilking$^{4}$, H.~J.~Yang$^{14}$,
        G.~P.~Zeller$^7$, E.~D.~Zimmerman$^{4}$ \\
\smallskip
(The MiniBooNE Collaboration)
\smallskip
}
\smallskip
\smallskip
\affiliation{
$^1$University of Alabama; Tuscaloosa, AL 35487 \\
$^2$Bucknell University; Lewisburg, PA 17837 \\
$^3$University of Cincinnati; Cincinnati, OH 45221\\
$^4$University of Colorado; Boulder, CO 80309 \\
$^5$Columbia University; New York, NY 10027 \\
$^6$Embry Riddle Aeronautical University; Prescott, AZ 86301 \\
$^7$Fermi National Accelerator Laboratory; Batavia, IL 60510 \\
$^8$University of Florida; Gainesville, FL 32611 \\
$^9$Indiana University; Bloomington, IN 47405 \\
$^{10}$Los Alamos National Laboratory; Los Alamos, NM 87545 \\
$^{11}$Louisiana State University; Baton Rouge, LA 70803 \\
$^{12}$Massachusetts Institute of Technology; Cambridge, MA 02139 \\
$^{13}$Instituto de Ciencias Nucleares, Universidad National Aut\'onoma de M\'exico, D.F. 04510, M\'exico \\
$^{14}$University of Michigan; Ann Arbor, MI 48109 \\
$^{15}$Princeton University; Princeton, NJ 08544 \\
$^{16}$Saint Mary's University of Minnesota; Winona, MN 55987 \\
$^{17}$Virginia Polytechnic Institute \& State University; Blacksburg, VA 24061 \\
$^{18}$Yale University; New Haven, CT 06520\\
}

\begin{abstract}
A high-statistics sample of charged-current muon neutrino scattering events 
collected with the MiniBooNE experiment is analyzed to extract the first measurement of the double
differential cross section ($\frac{d^2\sigma}{dT_\mu d\cos\theta_\mu}$) for 
charged-current quasielastic (CCQE) scattering on carbon. This result features minimal 
model dependence and provides the most complete information on this process to date. 
With the assumption of CCQE scattering, the absolute cross section 
as a function of neutrino energy ($\sigma[E_\nu]$) and the single differential cross section 
($\frac{d\sigma}{dQ^2}$) are extracted to facilitate comparison with previous measurements.  
These quantities may be used to characterize an effective axial-vector form factor of the nucleon 
and to improve the modeling of low-energy neutrino interactions on nuclear targets. The results
are relevant for experiments searching for neutrino oscillations.
\end{abstract}



\maketitle

\section{Introduction}\label{sec:intro}
Neutrino charged-current (CC) scattering without pions in the final state
is important to measure and characterize, and is a critical
component in the neutrino oscillation program of the MiniBooNE
experiment~\cite{MB_osc1,MB_osc2,MB_osc3,MB_osc4} at Fermilab. Most of these events 
are charged-current quasielastic scattering (CCQE) of 
the muon neutrino on a bound nucleon ($\nu_\mu + n \rightarrow \mu^- + p$).
A robust model of these interactions is required to support future experiments
such as NOvA~\cite{NOvA} and T2K~\cite{T2K} that are also searching for
$\nu_\mu \rightarrow \nu_e$ oscillations. Such experiments will use
$\nu_e$ CC interactions to
detect the appearance of any $\nu_e$ resulting from oscillations in the
large distance between production and detection. Additional use will
be made of $\nu_\mu$ CC interactions to normalize the neutrino content 
at production using a
near detector and to search for the disappearance of $\nu_\mu$ via a
far detector.  These analyses will require all available experimental
and theoretical insight on the CCQE interaction in the $\approx 1$~GeV
energy range and on nuclear (carbon, oxygen) targets. While many
unknown quantities are eliminated in these experiments by considering
ratios of far to near events, the cancellation is not complete due to
differences in neutrino flux and backgrounds in the near and far
detectors. Thus, in order to permit {\em precision} oscillation
measurements, it is important to have an accurate characterization 
of the CCQE differential cross sections over a wide span of neutrino energies.

Historically, it has proven difficult to accurately define the 
CCQE cross section and precise measurements have been unavailable.
The experimental 
execution and data interpretation are non-trivial for several reasons. 
Neutrino beams typically span a wide energy range thereby preventing an incoming 
energy constraint on the reaction. The neutrino flux itself is often poorly 
known, hampering normalization of reaction rates. Background 
processes are frequently significant and difficult or impossible to separate 
from the CCQE signal, for instance, CC pion production 
combined with pion absorption in the nucleus. Further complicating the 
description, the target nucleon is not free but bound in a nuclear target
and correlations between nucleons may be important.
There are differing detection strategies employed by different experiments,
for example, some require detection of the final state nucleon and some do 
not. Finally, the 
nuclear target often differs between experiments, thus making comparisons 
less straightforward.

The current data on CCQE scattering come from a variety of experiments 
operating at differing energies and with different nuclei~\cite{Zeller}. 
Modeling of this data has been consistent from experiment to experiment, yet
remains fairly unsophisticated. Preferred for its simplicity, neutrino CCQE models 
typically employ a relativistic Fermi gas (RFG) model (such as that of 
Ref.~\cite{SmithMoniz}) that combines the bare nucleon physics with a model
to account for the nucleon binding within the specific nucleus. 
The structure of the nucleon is parametrized with the 
three dominant form factors: two vector, $F_{1,2}(Q^2)$, and one axial-vector, 
$F_A(Q^2)$. The vector form factors, including the $Q^2$ (squared 
four-momentum transfer) dependence, are well-determined from electron scattering. The 
axial vector form factor at $Q^2=0$ is known from neutron beta-decay.
Neutrino-based CCQE measurements may then be interpreted as a measurement of 
the axial-vector mass, $M_A$, which 
controls the $Q^2$ dependence of $F_A$, and ultimately,
the normalization of the predicted cross section.  

This simple, underlying model has led to the situation where neutrino
CCQE measurements typically produce a measurement of $M_A$ independent of 
neutrino energy and target nucleus. The resulting world-average is 
$M_A=1.03\pm0.02$ GeV~\cite{past-ma} (a recent summary of the various $M_A$ 
values is provided in Ref.~\cite{NOMAD}). It should be noted that the data 
contributing to this world-average are dominated by higher precision 
bubble chamber experiments using deuterium as a target.  
In addition, most (but not all) of the $M_A$ values have come
from the observed distribution of CCQE events in $Q^2$ rather than from an
overall normalization of the event yield.

Several experiments have recently reported new results on CCQE scattering 
from high-statistics data samples with intense, well-understood neutrino beams. 
The NOMAD experiment extracted a CCQE cross section and 
$M_A$ from data taken on carbon in the energy range, 
$3<E_\nu<100$~GeV~\cite{NOMAD}. They observe an $M_A$ value and cross section 
consistent with the prior world-average. However, data at lower neutrino
energies using carbon or oxygen as a target have indicated, through 
$Q^2$-shape fits, a somewhat larger value for $M_A$ (by $\approx 
10-30$\%)~\cite{MB_CCQE,K2K_CCQE1,K2K_CCQE2}.  The SciBooNE 
experiment has recently reported a preliminary extraction of the 
total CCQE cross section on carbon that is consistent with these results~\cite{SB_CCQE}. 
To add to this, the MINOS experiment has reported a preliminary result on 
$M_A$ extracted from CCQE data collected on iron in a similar energy range 
as NOMAD~\cite{MarkD}. This result for $M_A$ is also larger than the world-average. 
None of these experiments has reported differential cross sections.

The CCQE cross section predictions resulting from these differing 
measured values for $M_A$ are too unreliable for use by future oscillation 
experiments, and the collection of existing results remains puzzling. 
Perhaps the currently employed model of the CCQE process is too naive and 
the physics of the bound nucleons can alter the extracted $M_A$. There is 
currently large theoretical interest in this area~\cite{new-model,Martini} and a 
solution may well emerge. Regardless, if the experimental results hold 
up, it is clear that improved measurements will be required to sort out the 
situation. These measurements will need to go beyond simply extracting an 
$M_A$ value as the data are already indicating that this strategy is too 
simplistic. Experiments should advance to providing cross sections, 
differential where possible, that are correctly normalized with a predicted 
neutrino flux (not normalized to a different reaction channel in the same data). 
Experiments should also strive to reduce the model-dependence of their reported 
results. To this end, we present differential cross sections in muon kinematics
extracted from $\nu_\mu$ CCQE events collected in MiniBooNE.

MiniBooNE has accumulated the world's largest sample of $\nu_\mu$ 
CCQE events  ($\approx 150,000$)  in the 1~GeV region. Using this high-statistics 
and low-background event sample, we report the
first measurement of an absolute $\nu_\mu$ CCQE double differential cross 
section, the main result of this work. In addition, CCQE cross sections in 
several other conventional forms are provided. The layout of 
the remainder of this paper is as follows. In Section~\ref{sec:mb}, 
we provide a summary of the MiniBooNE experiment, including the Booster 
Neutrino Beamline (BNB) and the MiniBooNE detector. We detail the neutrino 
interaction model used to describe the signal and background in 
Section~\ref{sec:nuan}. The CCQE selection and analysis strategy is outlined 
in Section~\ref{sec:analysis}. Finally, in Section~\ref{sec:xsec}, we report 
the MiniBooNE flux-integrated CCQE double differential cross section 
($\frac{d^2\si}{dT_\mu d\cos\th_\mu}$), the flux-integrated CCQE single 
differential cross section ($\frac{d\si}{dQ^2_{QE}}$), and the flux-unfolded 
CCQE cross section as a function of energy ($\si[E_\nu^{QE,RFG}]$). To 
facilitate comparison with updated model predictions~\cite{new-model,Martini}, we 
provide the predicted MiniBooNE neutrino fluxes and measured cross section 
values in tabular form in the Appendix.

\section{MiniBooNE experiment}\label{sec:mb}

\subsection{Neutrino beamline and flux}\label{sec:beam}
\begin{figure}
\includegraphics[width=\columnwidth]{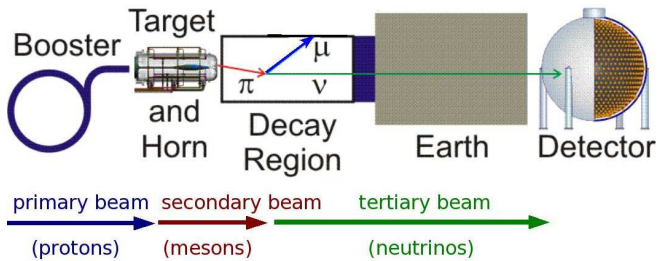}
\caption{(color online) Schematic overview of the MiniBooNE experiment 
including the Booster Neutrino Beamline and MiniBooNE detector.}
\label{fig:BNB}
\end{figure}

\begin{figure}
\includegraphics[width=\columnwidth]{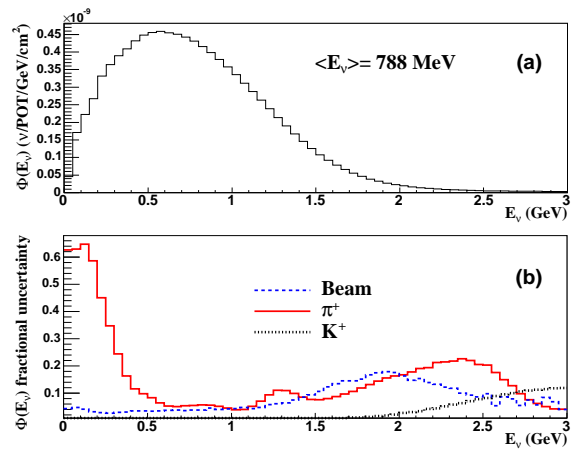} \\
\caption{(color online)
Predicted $\nu_\mu$ flux at the MiniBooNE detector (a) along with the 
fractional uncertainties grouped into various contributions (b).
The integrated flux is $\MBnumuflx$~$\numu$/POT/cm$^2$ 
($0<E_\nu<3$ GeV) with a mean energy of 788~MeV. Numerical values 
corresponding to the top plot are provided in Table~\ref{tab:nuflux} in 
the Appendix.}
\label{fig:nuflux} 
\end{figure}

The Booster Neutrino Beamline (BNB) consists of three major components as shown
in Figure~\ref{fig:BNB}: a primary proton beam, a secondary meson beam, and 
a tertiary neutrino beam. 
Protons are accelerated to 8~GeV kinetic energy in the Fermilab Booster 
synchrotron and then fast-extracted in 1.6~$\mu$s ``spills'' to the BNB. These 
primary protons impinge on a 1.75 interaction-length beryllium target centered
in a magnetic focusing horn. The secondary mesons that are produced are 
then focused by a toroidal magnetic field which serves to direct the resulting
beam of tertiary neutrinos towards the downstream detector. The neutrino flux 
is calculated at the detector with a \textsc{geant4}-based~\cite{GEANT4} 
simulation which takes into account proton transport to the target, interactions
of protons in the target, production of mesons in the $p$-Be process, and 
transport of resulting particles through the horn and decay volume. A full 
description of the calculation with associated uncertainties is provided in 
Ref.~\cite{beam}. MiniBooNE neutrino data is not used in any way to obtain the
flux prediction. The resulting $\nu_\mu$ flux is shown as a function of 
neutrino energy in Figure~\ref{fig:nuflux} along with its predicted 
uncertainty. These values are tabulated in Table~\ref{tab:nuflux} 
in the Appendix. The $\nu_\mu$ flux has an average energy (over $0<E_\nu<3$~GeV) 
of 788~MeV and comprises 93.6\% of the total flux of neutrinos at the
MiniBooNE detector. There is a 5.9\% (0.5\%) contamination of $\numub$
($\nue, \nueb$); all events from these (non-$\numu$) neutrino types
are treated as background in this measurement
(Section~\ref{sec:xsformula}).

The largest error on the predicted neutrino flux results from the
uncertainty of pion production in the initial $p-$Be process in the
target as the simulation predicts that 96.7\% of muon neutrinos in the
BNB are produced via $\pi^{+}$ decay. The meson production model in
the neutrino beam simulation~\cite{beam} relies on external hadron
production measurements. Those of the HARP experiment~\cite{HARP} are
the most relevant as they measure the $\pi^{\pm}$ differential cross
section in $p-$Be interactions at the same proton energy and on the
same target material as MiniBooNE.  The uncertainty in $\pi^{+}$
production is determined from spline fits to the HARP $\pi^{+}$
double differential cross section data~\cite{beam}.  The spline-fit
procedure more accurately quantifies the uncertainty in the underlying
data, removing unnecessary sources of error resulting from an
inadequate parameterization~\cite{Sanford-Wang} of the HARP data. The
HARP data used was that from a thin (5\% interaction length) beryllium
target run~\cite{HARP}. While that data provides a valuable constraint
on the BNB flux prediction, additional uncertainties resulting from
thick target effects (secondary re-scattering of protons and pions)
are included through the BNB flux simulation.

The resulting $\pi^{+}$ production uncertainty is $\approx$ 5\% at the peak 
of the flux distribution increasing significantly at high and low neutrino 
energies. There is a small contribution to the $\numu$ flux error from the 
uncertainty in kaon production which is significant only for $E_\nu > 2.0$~GeV.
Other major contributions to the flux error include uncertainties on the 
number of protons on target (POT), hadron interactions in the target, and the
horn magnetic field. These are grouped as the ``beam'' component shown 
together with the aforementioned pion and kaon production uncertainties in 
Figure~\ref{fig:nuflux}b. All flux errors are modeled through variations in the
simulation and result in a total error of $\approx$ 9\% at the peak of the 
flux. In practice, a complete error matrix is calculated in bins of neutrino 
energy that includes correlations between bins. This matrix is used to 
propagate the flux uncertainties to the final quantities used to extract 
the cross section results reported here.

\subsection{MiniBooNE detector}\label{sec:dtec}
The MiniBooNE detector (shown schematically in Figure~\ref{fig:BNB}) is
located 541~m downstream of the neutrino production target and
consists of a spherical steel tank of 610~cm inner radius filled
with 818 tons of Marcol7 light mineral oil (CH$_2$) with a density of
0.845~g/cm$^3$.  The volume of the tank is separated into an inner and
an outer region via an optical barrier located at a radius of 574.6~cm.  The
inner and outer regions are only separated optically, the oil is the
same in each.  The index of refraction of the oil is 1.47, yielding a
Cherenkov threshold for particles with $\beta>0.68$. The mineral
oil is undoped, that is, no additional scintillation solutes were
added.  However, because of intrinsic impurities in the oil, it produces 
a small amount of scintillation light in addition to Cherenkov light 
in response to energy loss by charged particles.
 
The inner (signal) region is viewed by an array of 1280 inward-facing 8-inch
photo multiplier tubes (PMTs) mounted on the inside of the optical
barrier and providing 11.3\% photocathode coverage of the surface 
of the inner tank region.  The outer (veto) region is monitored by 240
pair-mounted PMTs which record the light produced by charged particles
entering or exiting the detector volume.

The PMT signals, in response to the light produced from charged particles, are
routed to custom-built electronics modules where they are amplified,
discriminated, and digitized.  These (``QT'') modules extract the
start time and integrated charge from each PMT pulse that is above
a discriminator threshold of $\approx$0.1~photoelectron.
This unit of data is called a PMT ``hit''.  The data is stored in a temporary
buffer until a trigger decision is made.  The trigger system uses
information from the Fermilab accelerator clock signals and PMT
multiplicities to form physics and calibration trigger signals. The
physics trigger for this analysis requires only that beam be
sent to the BNB neutrino production target.  When this condition is
satisfied, all PMT-hit data in a 19.2~$\mu$s window starting 5$~\mu$s before
the 1.6~$\mu$s beam spill is extracted from the QT modules and
added to the data stream.  This readout strategy collects all PMT data 
(with no multiplicity threshold) from beam-induced neutrino events
as well as any muon decays that occur with a characteristic time of 2~$\mu$s 
after the neutrino interaction.

The data within the 19.2~$\mu$s readout window is examined at
the analysis stage to organize the hits into temporal clusters or
``subevents''.  A subevent is any group of at least 10 hits (from
inner or outer PMTs) where no 
two consecutive hits are separated in time by more than 10~ns. These
subevents may then be analyzed separately to extract further information 
such as energy and position. With this scheme, muon-decay electrons 
or positrons may be identified and separated from the primary neutrino
event.

The MiniBooNE detector is calibrated via the light from a pulsed laser
source, cosmic muons, and muon-decay electrons.  The laser
calibration system consists of a pulsed diode laser injecting light
via optical fibers into four 10~cm dispersion flasks located at
various depths in the main detector volume. In addition,
one bare fiber is installed at the top of the tank.  This system is
used to quantify the charge and time response of the PMTs and allows
for an {\em in situ} measurement of the oil attenuation length and
light scattering properties. 

The muon calibration system consists of a two-layer scintillation
hodoscope located above the detector tank combined with seven 5-cm
cubic scintillators deployed at various locations within the tank (near 
the vertical axis) and used to tag stopped muons. With this system, stopping
cosmic ray muons of energies ranging from 20 to 800~MeV are tracked through the
detector enabling an energy calibration via the known range of the stopping muons. 
Using this data, the energy (angular) resolution for reconstructed muons in MiniBooNE is 
measured to be 12\% ($5.4^\circ$) at 100~MeV, improving to 3.4\% ($1.0^\circ$) 
at 800~MeV.

A separate large sample of stopping cosmic muons is obtained via a
dedicated calibration trigger that requires the signature of an
incoming muon and its decay electron.  This sample allows for the
calibration and measurement of the detector response to muon-decay
electrons.

More details about the oil medium, detector structure, PMTs, electronics,
and calibration are available in Ref.~\cite{dtec}.

\subsection{Detector simulation}\label{sec:detsim}
A \textsc{geant3}-based~\cite{GEANT3} program is used to simulate the response
of the detector to neutrino interactions.  
This simulation is used to determine the detection efficiency 
for CCQE events, the probability for accepting background events, and
the error on relevant observables due to uncertainties in the detector
response.

The entire geometry of the MiniBooNE detector is modeled including the
detector tank and all inner components. The major components of the
detector housing structure are modeled as is the surrounding
environment. Standard \textsc{geant3} particle propagation and decay routines are
utilized with some changes made to better simulate 
$\pi^0$ decay, $\mu$ decay, and $\mu^-$ capture on carbon.  The latter process is
especially important for the background estimation in the analysis
reported here.  The \textsc{gcalor}~\cite{GCALOR} hadronic interaction package
is used in place of the default \textsc{gfluka} package in \textsc{geant3} 
because it better reproduces known data on $\pip$ absorption and 
charge exchange in the relevant $\pip$ energy range (100-500~MeV).  

Some modifications were made to the standard \textsc{GCALOR} code to better
simulate $\pi^\pm$ processes.  Pion radiative capture/decay and photonuclear 
processes are of concern for the neutrino oscillation search~\cite{MB_osc2}
but have negligible effect for this analysis. Elastic scattering of $\pi^\pm$ 
on carbon is important  and was not simulated in the 
standard \textsc{GCALOR} code.  A model, guided by the available 
data~\cite{Ashery,Binon}, was added to the simulation and yields only
a small change to the calculated background from pion production processes.
   
Uncertainties of 35\% on $\pip$ absorption and 50\% on $\pip$ charge exchange
are assigned based on the difference between the external data~\cite{Ashery} 
and the \textsc{GCALOR} prediction. Note that these errors are relevant for $\pip$ 
propagation in the {\em detector medium} not {\em intranuclear} processes which 
are assigned separate uncertainties (Section~\ref{sec:nufsi}).

Charged particles propagating through the detector oil produce optical
photons via Cherenkov radiation and scintillation.  Optical wavelengths
of 250-650~nm are treated.  The Cherenkov process is
modeled with standard \textsc{geant3}.  The scintillation process
is modeled with a MiniBooNE-specific simulation that creates optical
photons at a rate proportional to Birk's law-corrected energy loss
with an emission spectrum determined from dedicated florescence
measurements.  Optical photons resulting from these production
processes are tracked through the detector oil with consideration of
scattering, fluorescence, absorption, and reflection from detector
surfaces (including PMTs). Photons that intersect the PMT surface (and
do not reflect) are modeled with a wavelength and incident
angle-dependent efficiency. The photon signal in each PMT is used
together with the known response of the PMT and readout electronics to
generate simulated data that is then input to the data analysis programs.

The models, associated parameters, and errors implemented in the detector
simulation are determined through external measurements
of the properties of materials as well as internal measurements using data
collected in the MiniBooNE detector.  It is particularly important to
correctly model the optical photon transport since a typical optical photon
travels several meters before detection.  This ``optical model'' is
tuned starting from various external (to the detector) measurements of
MiniBooNE mineral oil optical properties, such as the refractive index,
attenuation length, and fluorescence/scintillation strength.
These quantities allow for the implementation of various
models to describe the optical photon propagation.  The details of the
models are then further adjusted based on MiniBooNE internal
data sets such as muon-decay electrons, cosmic muons, and laser
pulses.  In total, 35 optical model parameters are adjusted to
obtain a good description of the various data sets.  Values for the
uncertainties of these parameters, including correlations among them,
are also extracted from the data and the effect on the reported
observables determined by running the simulation with adjusted
values (Section~\ref{sec:errmat}).

Additional details about the MiniBooNE detector simulation and supporting
measurements of oil properties are available in Refs.~\cite{dtec,Ryan}.
 
\section{Neutrino interaction model}\label{sec:nuan}
The MiniBooNE experiment employs the \nuance\ \textsc{v3} event 
generator~\cite{nuance} to estimate neutrino interaction rates in the CH$_2$ 
target medium.  The \nuance\ generator considers all 
interaction processes possible in the neutrino energy region relevant for
MiniBooNE.  It also enables the various processes to be tuned to match
the data via input parameters or source code changes where necessary.  
The \nuance\ generator includes the following components: 
(1) a relativistic Fermi gas (RFG) model for CCQE (and NC elastic) scattering 
    on carbon~\cite{SmithMoniz}; 
(2) a baryonic resonance model for CC/NC single and multi-pion 
    production~\cite{ReinSehgal_res};  
(3) a coherent CC/NC single pion production model~\cite{ReinSehgal_coh};
(4) a deep inelastic scattering (DIS) model~\cite{GRV98,BodekYang};  and 
(5) a final-state interaction model to simulate re-interaction of final state 
hadrons in the nuclear medium~\cite{nuance}. Neutrino interactions on both free 
(protons) and bound nucleons (in carbon) are considered to model the
CH$_2$ detector medium. Further details on these models, parameters, and 
uncertainties are provided in the following subsections.

This event generator is used, after the adjustment of parameters to adequately 
describe the MiniBooNE data, to calculate the background contribution to the 
CCQE signal. The CCQE model is used only in the extraction of the model
parameters in a shape-fit to the $Q^2_{QE}$ distribution 
(Section~\ref{sec:ccpifit}).  The CCQE differential cross section measurements 
do not depend on this model (excepting some small dependence due to detector 
resolution corrections and $\numub$ CCQE backgrounds).
In addition, since it is such a large background to 
CCQE, the CC1$\pip$ background is constrained (outside of the \nuance\ model) 
to reproduce MiniBooNE data (Section~\ref{sec:ccpifit}). A summary of 
interaction channels considered and the \nuance\ predictions for the neutrino 
interaction fractions in the MiniBooNE detector are provided in 
Table~\ref{tab:pichart}. The final values for the predicted backgrounds 
(after event selection cuts) are provided in Table~\ref{tab:CCQEevsum}.

\begin{table*}
\begin{tabular}{p{2.0in}p{1.in}lr}
\hline
\hline
neutrino process                  & abbreviation              & reaction                               & fraction (\%)\\
\hline
CC quasielastic                   & CCQE                      &$\nu_\mu+n\to\mu^-+p$                       &$\fCCQEt$\\
NC elastic                        & NCE                       &$\nu_\mu+p(n)\to\nu_\mu+p(n)$               &$\fNCELt$\\
CC 1$\pip$ production             & CC1$\pip$                 &$\nu_\mu+p(n)\to\mu^-+\pip+p(n)$           &~~$\fCCPIt$\\
CC 1$\piz$ production             & CC1$\piz$                 &$\nu_\mu+n\to\mu^-+\piz+p$                  &~~$\fCCPZt$\\
NC 1$\pi^\pm$ production          & NC1$\pi^\pm$              &$\nu_\mu+p(n)\to\nu_\mu+\pip(\pim)+n(p)$     &~~$\fNCPPt$\\
NC 1$\piz$ production             & NC1$\piz$                 &$\nu_\mu+p(n)\to\nu_\mu+\piz+p(n)$            &~~$\fNCPZt$\\
multi pion production, DIS, etc.  & other                     &$\nu_\mu+p(n)\to\mu^{-}+N\pi^\pm+X$, etc. &~~$\fOTHRt$\\

\hline
\hline
\end{tabular}
\caption{
Event type nomenclature and \nuance-predicted $\nu_\mu$ event fractions for 
MiniBooNE integrated over the predicted flux in neutrino mode before 
selection cuts. For the pion production channels, indirect production (through resonance states)  
and direct production (through coherent processes) are included.
(CC=charged-current, NC=neutral-current).}
\label{tab:pichart}
\end{table*}

\subsection{Charged current quasielastic scattering}\label{sec:nuccqe}
CCQE scattering is the dominant neutrino interaction process in MiniBooNE
and the subject of this analysis.  This process is defined as the charge-changing
scattering of a neutrino from a single nucleon with no other particles
produced and it is simulated with the RFG 
model~\cite{SmithMoniz} with several modifications. 
A dipole form is used for the axial form factor
with an adjustable axial mass, $M_A$. An empirical Pauli-blocking parameter, $\ka$,
is introduced~\cite{MB_CCQE} to allow for an extra degree of freedom that is 
important to describe the MiniBooNE data at low momentum transfer. This parameter
is a simple scaling of the lower bound, $\Elo$, of the nucleon energy integral of
Ref.~\cite{SmithMoniz}
via $\Elo=\kappa(\sqrt{p_F^2+M_p^2} - \omega+E_B)$, where $p_F$, $M_p$, $\omega$, and $E_B$
are the Fermi momentum, nucleon mass, energy transfer, and binding energy, respectively.
When $\kappa>1$, the Pauli-blocking of final-state
nucleons is increased which reduces the cross section at low momentum transfer.

A parametrization~\cite{BBA03} is used to describe the non-dipole behavior of the Dirac 
and Pauli form factors. Although the contribution is small, the 
pseudoscalar form factor, derived from partial conservation of the axial 
vector current (PCAC) is also included~\cite{fp}. The scalar and axial tensor 
form factors are set to zero as implied from G-parity conservation. The Fermi 
momentum and binding energy for carbon are set to $220\pm30$~MeV/c and 
$34\pm9$~MeV, respectively, as extracted from electron scattering 
data~\cite{Moniz} taking account of the purely isovector character of CCQE.

The parameters $M_A$ and $\ka$ were initially extracted from MiniBooNE
CCQE data in a prior analysis~\cite{MB_CCQE}, and 
were determined to be $\MAeff=1.23\pm0.20$~GeV and $\ka=1.019\pm0.011$.
While not the main result of this paper, this exercise is repeated
after explicitly measuring the CC1$\pip$ background from MiniBooNE data 
and is described in Sect~\ref{sec:ccpifit}. The superscript ``$\mathrm{eff}$''
on $\MA$ was introduced to allow for the possibility that the axial mass 
measured from scattering on nucleons bound in carbon may be different from 
the ``bare-nucleon'' axial mass that appears within the neutrino model.  The 
use of this notation is continued in this work. The uncertainties in these 
CCQE model parameters do not propagate to the errors on the measured cross 
sections for this channel.  
Neutral current elastic (NCE) scattering is described with the same model as 
that for the CCQE interaction with the replacement of appropriate form 
factors~\cite{BNL734_garvey} to describe the NC coupling to the nucleon.
The uncertainty from the NCE model parameters on the CCQE results
is negligible due to the small background contribution from this
channel (Table~\ref{tab:CCQEevsum}).

\subsection{Resonance interactions}\label{sec:nures}
The primary source of single pion production for MiniBooNE is predicted
to be baryonic resonance production and decay, such as, 
\begin{align*}
\nu_\mu+p\to\mu^-+&\De^{++}         \\
                  &\hookrightarrow \pip+~p,\\
\nu_\mu+n\to\mu^-+&\De^{+}          \\
                  &\hookrightarrow \pip+n~, \piz+p. 
\end{align*}
The \nuance\ model employs the relativistic harmonic oscillator quark model of 
Ref.~\cite{ReinSehgal_res,FKR}. The pion angular distribution due to the spin 
structure of the resonance states is additionally taken into account.  
In total, 18 resonances and their interferences are simulated in reactions with
invariant mass $W<2$~GeV, however, the $\De (1232)$ resonance dominates at 
this energy scale. For reactions on bound nucleons, an RFG model is employed 
with a uniform Fermi momentum and constant binding energy. In-medium effects 
on the width of resonances are not considered explicitly, however, final-state
interactions can produce an effective change in these widths. An axial mass of 
$M_A^{1\pi}=1.10\pm0.27$~GeV, set by tuning to available data, is used for this channel.

Multi-pion processes are considered in the \nuance\ simulation with 
$M_A^{N\pi}=1.30\pm0.52$~GeV. This parameter was set (together with $M_A^{1\pi}$) so 
that the simulation reproduces inclusive CC data.  The contribution of 
this channel to the CCQE background is small and the uncertainty is negligible 
in the final errors.

The CC1$\pip$ channel is the largest background contribution to the CCQE 
signal and the uncertainty from the model prediction alone is 
substantial~\cite{JarekN}. However, the experimental signature of the 
CC1$\pip$ reaction in MiniBooNE is distinct and the efficiency is large.  So, 
in order to reduce uncertainty stemming from the CC1$\pip$ model prediction,
a measurement of the CC1$\pip$ background is performed as part of the CCQE 
analysis (Section~\ref{sec:ccpifit}).

\subsection{Coherent pion production}\label{sec:nucohpi}
Pions are also produced in the CC and NC coherent interaction of neutrinos 
with carbon nuclei, 
\begin{align*}
\nu_\mu+A&\to \mu^-+\pip+A~,  \\
\nu_\mu+A&\to \nu_\mu+\piz+A~.
\end{align*}
In \nuance, this coherent pion production process is described using the 
model of Ref.~\cite{ReinSehgal_coh} assuming 
$M_A^{coh}=1.03\pm0.28$~GeV~\cite{nuance}.

Coherent scattering is predicted to have distinct features in the angular 
distributions of both the final-state muons and pions. Both the
K2K~\cite{K2K_cohpi} and MiniBooNE~\cite{MB_NCpi0} experiments have measured 
the fraction of pions produced coherently in $\approx$~1 GeV neutrino 
interactions. K2K measured a rate for coherent 
CC1$\pip$ production consistent with zero and set an upper limit. MiniBooNE 
measured a non-zero value for coherent NC1$\piz$ production albeit $\approx$~35\% 
smaller than the model prediction~\cite{nuance,ReinSehgal_coh}. The latest 
result from the SciBooNE experiment is consistent with the K2K measurement 
for CC1$\pip$ coherent production~\cite{SB_cohpi}. 

Because of the current discrepancy between CC and NC coherent pion 
measurements and the variation in model predictions at low energy, the
prediction is reduced to 50\% of the default value in 
\nuance~\cite{nuance,ReinSehgal_coh} and assigned a 100\% uncertainty.  
This choice spans the current results from relevant experiments and existing 
theoretical predictions.

\subsection{Final State Interactions}\label{sec:nufsi}
In the \nuance\ simulation, neutrino interactions on nucleons are 
modeled using the impulse approximation which assumes the interaction occurs 
instantaneously on independent nucleons. The binding of nucleons within 
carbon is treated within the RFG model, however, any nucleon-nucleon correlation
effects are not. The final state hadrons may interact within the nucleus
as they exit. They are propagated through the $^{12}$C nucleus with a known, radially-dependent 
nucleon density distribution~\cite{Carbonden} and may undergo final-state 
interactions (FSI). These are simulated by calculating interaction probabilities
for the possible processes in 0.3~fm steps until the particles leave the 
$\approx$~2.5~fm-radius spherical carbon atom. The interaction probabilities are 
derived from external $\pi-N$, $N-N$ cross section and angular distribution 
data~\cite{PiN}, as well as the nuclear density of carbon. 

To model $\De$ absorption in the nucleus, ($\De+N\to N+N$), a constant, energy-independent
probability for an intranuclear interaction of 20\% (10\%) is 
assumed for $\De^++N$, $\De^{0}+N$
($\De^{++}+N$, $\De^{-}+N$) processes. These values were chosen based 
on comparisons to K2K data~\cite{Casper} and are assigned a 100\% uncertainty.
After an interaction, the density distribution and step
size are modified to prevent an overestimate of these FSI 
effects~\cite{Carbonmodi,Hawker}. 

Of these hadron FSI processes, intranuclear pion absorption and pion charge 
exchange, ($\pip+X \to X', \pip+X \to \piz+X'$)  are the most important 
contributions to the uncertainty in the CCQE analysis.  Pion absorption and 
charge exchange in the detector medium are addressed separately in the 
detector simulation (Section~\ref{sec:detsim}).

\begin{figure}
\begin{center}
\includegraphics[width=\columnwidth]{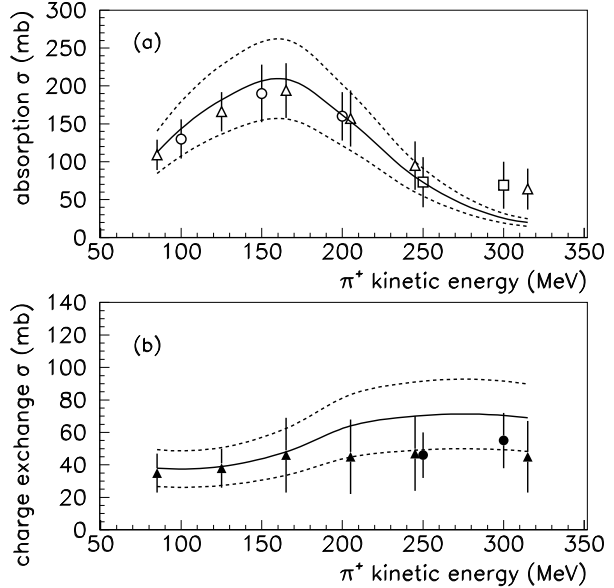}
\end{center}
\caption{
A comparison of relevant data~\cite{Ashery} with the  
\nuance\ model (solid lines) for intranuclear pion (a) absorption and (b) 
charge exchange as a function of pion kinetic energy. The dotted 
lines show the 25\% (30\%) systematic error bands assumed for the pion
absorption (charge-exchange) cross sections. }
\label{fig:pion-abs-ce}
\end{figure}

A CC1$\pip$ interaction followed by intranuclear pion absorption 
is effectively indistinguishable from the CCQE process in MiniBooNE
(they are ``CCQE-like'', Section~\ref{sec:xsformula}).
An event with intranuclear pion charge exchange is distinguishable in the
detector, albeit not with 100\% efficiency.
These effects, combined with the high rate of CC1$\pip$ events, results in a 
significant background to the CCQE measurement that must be treated carefully. 
The model for these intranuclear pion processes has been tuned to match
the available data~\cite{Ashery} in the relevant pion energy
range.  A comparison of the adjusted \nuance\ model and relevant 
data is shown in Figure~\ref{fig:pion-abs-ce}.  A 25\% (30\%) systematic
error in the overall interaction cross section is used for the pion absorption 
(charge-exchange) process.

\section{CCQE measurement}\label{sec:analysis}
The goal of this measurement is to determine the double differential cross 
section for the CCQE process on carbon, $\nu_\mu+n\to\mu^-+p$, where the 
target neutron is bound in $^{12}$C.

\begin{figure}
\begin{center}
\includegraphics[width=6cm]{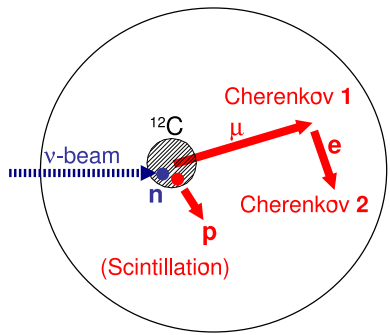}
\end{center}
\caption{
(color online) 
Schematic illustration of a CCQE interaction in the MiniBooNE detector. 
The primary Cherenkov light from the muon (Cherenkov 1, first subevent) 
and subsequent Cherenkov light from the decay-electron 
(Cherenkov 2, second subevent) are used to tag the CCQE event. No 
requirements are made on the outgoing proton.}
\label{fig:interaction}
\end{figure}

The identification of CCQE interactions in the MiniBooNE detector 
relies solely on the detection of the Cherenkov light from the primary 
(prompt) muon and the associated decay-electron. An illustration of this 
process is shown in Figure~\ref{fig:interaction}.  Scintillation light is
produced by the charged lepton and the recoil proton (or nuclear fragments).
However, with the reconstruction employed here, this light is not separable from the 
dominant Cherenkov light.  In addition, the proton is typically below Cherenkov
threshold.  These conditions are such that the proton is not separable from
the charged lepton and so no requirement is placed on the recoil 
proton in this analysis. This is to be contrasted with some measurements 
of CCQE interactions that do require the observation of a recoil 
proton for some part of the event sample~\cite{NOMAD,K2K_CCQE1,K2K_CCQE2,SB_CCQE}. 
An advantage of this insensitivity to the proton recoil is that the extracted
cross sections are less dependent on proton final-state model uncertainties.
However, the disadvantage in not detecting the recoil nucleon is that 
contributions to scattering from other nuclear configurations (such as 
two-nucleon correlations) are inseparable.  These contributions are, 
in the strictest sense, not CCQE, but counted as such in our experimental
definition.

A requirement of low veto activity for the CCQE sample 
ensures that all particles produced in the event stop
in the main region of the detector. This allows muons to be tagged with
high efficiency via their characteristic electron-decay with 
$\tau\approx 2$~$\mu$s.  

The CCQE interaction, including the muon decay, proceeds as,
\[
\begin{array}{cccl}
1: & \numu + n  & \to &  \mum + p \\ 
2: &            &     &  \hookrightarrow \elm + \nueb + \numu. 
\end{array} 
\]
 where each line in this equation identifies a subevent 
(Section~\ref{sec:dtec}). The primary muon is identified 
with the first subevent and
the subsequent decay-electron with the second subevent.
At BNB neutrino energies, neutrino interaction events that contain a 
primary muon predominantly result from CCQE scattering as can be seen in 
Table~\ref{tab:pichart}.  

The largest background is from CC single-pion production (CC1$\pip$). A
CC1$\pip$ interaction in the detector consists of (with subevents labeled), 
\[
\begin{array}{cccl}
&1:&  \hspace{-1.0in} \numu +p(n)  \to  \mum + p(n) + \pip  \\
&  &          \hspace{1.2in} \hookrightarrow  \mup + \numu \\
&2:&          \hspace{-0.1in} \hookrightarrow \elm + \nueb + \numu  \\
&3:&          \hspace{2.0in} \hookrightarrow \elp + \nue  + \numub. 
\end{array} 
\]
Note that this interaction results in three subevents: the primary
interaction and two muon decays (the muon decays can occur in any order). 
The $\pip$ decays immediately and
light from the prompt decay products contribute to the total light
in the primary event. 
These events may be removed from the CCQE sample by requiring exactly
two subevents. This requirement also reduces the
background from NC processes to an almost negligible level
because they do not contain muons and thus have only one subevent. This simple strategy 
results in a fairly pure sample of CCQE events.  
However, a significant number of CC1$\pip$ events have only two
subevents because one of the decay electrons escapes detection:
the $\mu^-$ is captured on $^{12}$C in the mineral oil (with 8\% probability~\cite{muoncap})
or the $\pip$ is absorbed.
Additionally, the study of CC1$\pip$ events for this analysis has indicated 
that the prediction for the CC1$\pip$ channel from the \nuance\ event generator 
is not sufficiently accurate for this measurement~\cite{JarekN}.  
For these reasons, the CC1$\pip$ rate is measured using a dedicated event sample. 
This differs from our previous strategy~\cite{MB_CCQE} 
where the default \nuance-predicted CC1$\pip$ fraction (with no adjustments) 
was used, with generous errors, in fits to the CCQE sample. 

The resulting procedure for selecting the CCQE sample and measuring the 
CC1$\pip$ background involves the following steps:
\begin{enumerate}
\item selection of a ``super-sample'' of events with a clean muon signature to 
isolate CC events (predominantly CCQE and CC1$\pip$) via analysis cuts;
\item application of a subevent cut to separate the super-sample into 
      CCQE (2-subevents) and CC1$\pip$ (3-subevents) samples;
\item measurement of the CC1$\pip$ rate from  the CC1$\pip$ sample;
\item adjustment of the CC1$\pip$ model in the event simulation to reproduce 
the measured rate; and 
\item subtraction of  this adjusted CC1$\pip$ background (along with other 
predicted backgrounds) from the CCQE signal to produce a a measurement of 
the CCQE interaction cross section. 
\end{enumerate}
The details of this procedure are provided in the following subsections.

In this analysis, the reconstruction of the CC1$\pip$ sample is for the sole 
purpose of background estimation.  Dedicated measurements of the CC1$\pip$ 
and  CC1$\piz$ channels in MiniBooNE have been reported 
elsewhere~\cite{CCpi+ratio,MikeW,BobN} including 
detailed reconstruction of the $\pip$ and $\piz$ kinematics.

\subsection{Event reconstruction}\label{sec:evt_recon}
For this analysis, it is crucial to identify and measure the muon in
the CC interaction. This is accomplished with an 
``extended-track'' reconstruction algorithm~\cite{recon} which uses
the charge and time information from all PMT hits in the first subevent
to form a likelihood that is maximized to determine the best single track 
hypothesis quantified by the 
track starting point, starting time, direction, and kinetic energy. This is 
performed with both a muon and electron particle hypothesis from which a (log) 
likelihood ratio is formed to enable particle identification.

The muon kinetic energy, $T_\mu$, and muon scattering angle, $\th_\mu$,
are extracted from the track reconstruction assuming a muon hypothesis.
These are used to form the fundamental observable reported here, the 
double-differential cross section.  For additional reported observables,
the reconstructed neutrino energy $E_\nu^{QE}$ and 
reconstructed four-momentum transfer $Q^2_{QE}$ are obtained via,
\beq
E_\nu^{QE}
&=&
\fr{2(M_n^{\prime})E_\mu-((M_n^{\prime})^2+m_\mu^2-M_p^2)}
{2\cdot[(M_n^{\prime})-E_\mu+\sqrt{E_\mu^2-m_\mu^2}\cos\th_\mu]},\label{eq:recEnu}\\
Q^2_{QE}
&=&
-m_\mu^2+2E_\nu^{QE}(E_\mu-\sqrt{E_\mu^2-m_\mu^2}\cos\th_\mu),\label{eq:recQsq}
\eeq
where $E_\mu = T_\mu + m_\mu$ is the total muon energy and $M_n$, $M_p$, $m_\mu$
are the neutron, proton, and muon masses. The adjusted neutron mass, $M_n^{\prime}=M_n-E_B$,
depends on the binding energy (or more carefully stated, the separation energy) in carbon, 
$E_B$, which for this analysis is set to $34\pm9$~MeV.

The subscript, ``$QE$'', on these reconstructed quantities is to 
call attention to these specific definitions and to distinguish them from 
quantities obtained
in other ways such as fits to the underlying true kinematic quantities.  
These are kinematic definitions that assume the initial nucleon (neutron) is at
rest and the interaction is CCQE (``QE assumption'').  While these quantities
certainly differ from the underlying true quantities, they are
well-defined, unambiguous, and easily reproduced by others. 

\subsection{CCQE and CC1$\pip$ event selection}\label{sec:evnt}
The CCQE and CC1$\pip$ candidate events are selected for this analysis 
and separated with a sequence of cuts summarized 
in Table~\ref{tab:cut}.  

\begin{table*}
\begin{tabular}{lp{0.6\textwidth}cc}
\hline
\hline
       &                                                             &\multicolumn{2}{c}{CCQE}   \\
cut \# & \multicolumn{1}{c}{description}                             &effic.(\%)  & purity(\%) \\
\hline
1     & all subevents, \# of veto hits $<6$                         &$\vetoCCQEeff$ & $\vetoCCQEpur$ \\
2     & 1st subevent, event time window, $4400<T(\mathrm{ns})<6400$ &$\timeCCQEeff$ & $\timeCCQEpur$ \\
3     & 1st subevent, reconstructed vertex radius $<500$~cm         &$\radiCCQEeff$ & $\radiCCQEpur$ \\
4     & 1st subevent, kinetic energy $>200$~MeV                     &$\enerCCQEeff$ & $\enerCCQEpur$ \\
5     & 1st subevent, $\mu/e$ log-likelihood ratio $>0.0$           &$\lemuCCQEeff$ & $\lemuCCQEpur$ \\
6     & \# total subevents $=2$ for CCQE ($=3$ for CC1$\pip$)       &$\tsubCCQEeff$ & $\tsubCCQEpur$ \\
7     & (CCQE-only) 1st subevent, $\mu-e$ vertex distance $>100$~cm and &                          \\
      & $\mu-e$ vertex distance $>(500 \times T_\mu(\mathrm{GeV})-100$)~cm &$\distCCQEeff$ & $\distCCQEpur$ \\
\hline
\hline
\end{tabular}
\caption{List of cuts for the CCQE and CC1$\pip$ event selections. The
predicted efficiency and purity values are for the CCQE signal normalized to all
CCQE events with a reconstructed vertex radius, $r<550$~cm.}
\label{tab:cut}
\end{table*}

The first five cuts are designed to efficiently select a high-purity 
sample of CCQE and  CC1$\pip$ events.
Cut~1 rejects events with incoming particles such as cosmic rays
or neutrino-induced events produced in the surrounding material.  It
also eliminates events where any of the neutrino interaction products
escape the main detector volume.  This is important for an accurate muon 
energy measurement and to avoid missing muon decays which leads to higher
backgrounds. Cut~1 does reduce the efficiency substantially (Tab.~\ref{tab:cut}),
however, it is necessary to reduce background (together with the subsequent
cuts). Cut~2 requires that the primary (muon) is in-time with the BNB spill window.
Cut~3 ensures that the reconstructed primary muon vertex is located
within a fiducial region in the main detector volume sufficiently far from
the PMTs for accurate reconstruction. Cut~4 
provides a minimum muon kinetic energy for reliable
reconstruction and reduces backgrounds from beam-unrelated muon-decay electrons.

\begin{figure}
\includegraphics[width=\columnwidth]{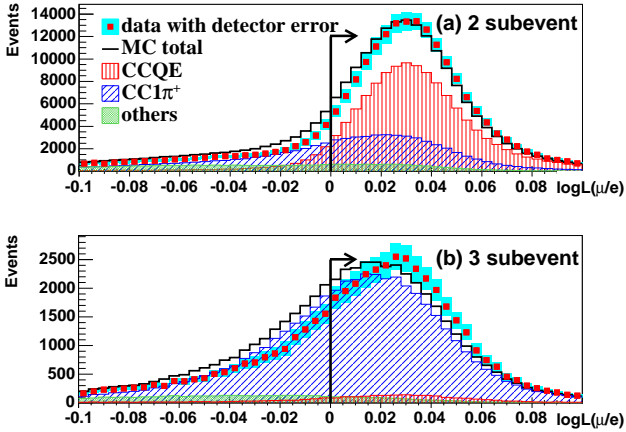}
\caption{
(color online)
Distributions of the $\mu/e$ log-likelihood ratio for the (a) 2 and (b) 3 
subevent samples. Data and Monte Carlo simulation (MC) are shown along with 
the individual MC contributions from CCQE, CC1$\pip$, and other channels. 
The lines with arrows indicate events selected by the muon-electron likelihood 
ratio cut.}
\label{fig:lemu}
\end{figure}

Cut~5 requires that the candidate primary muon is better fit as a muon
than as an electron.  Misreconstructed and multi-particle 
events tend to prefer the electron hypothesis so this cut reduces such
contamination.  This also substantially reduces the efficiency for selecting 
CC1$\pip$ events as can be seen in Figure~\ref{fig:lemu} where 
the $\mu/e$ log-likelihood ratio distribution is shown for each of the 2- and 3-subevent 
samples.  This bias is intended as it selects a sample of CC1$\pip$ with muon 
kinematics more closely matched to those CC1$\pip$ that are background
to the CCQE sample.  As is shown in  Fig.~\ref{fig:lemu}, data and Monte Carlo 
simulation (MC) agree fairly well to within the detector errors.   
The log-likelihood ratio distribution is quite sensitive to details of an 
event such as scintillation from hadron recoil via the PMT charge and time 
information~\cite{recon}.  The data-MC difference in the number of events
passing Cut~5 in both the 2- and 3-subevent samples is covered by the
full systematic errors considered in this analysis. 

Cut~6 separates the samples into CCQE (2 subevents) and CC1$\pip$ (3
subevents) candidates.  For this analysis, the second and third
subevents are required to contain at least 20 tank hits to reduce the
probability of accidental coincidences with the initial neutrino
interaction (first subevent).  This requirement reduces the efficiency for
identifying the muon-decay electron by $\approx 3\%$.

\begin{figure}
\includegraphics[width=\columnwidth]{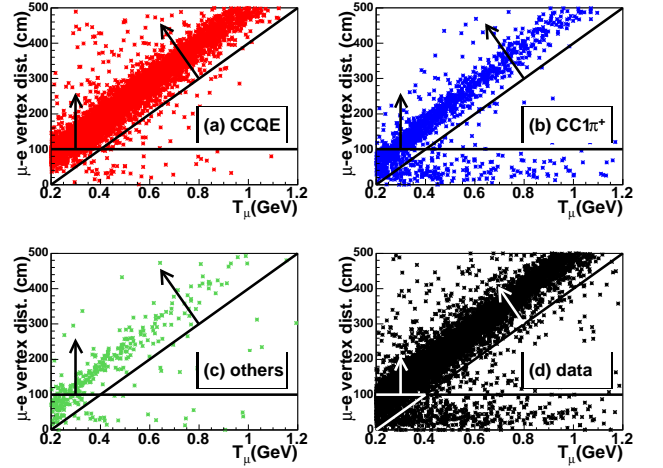}
\caption{
(color online)
Scatter plots of muon-electron vertex distance as a function of reconstructed
muon kinetic energy for MC samples: (a) CCQE, (b) CC1$\pip$, and (c) other 
channels. The distribution of data is shown in (d). These are events with 2-subevents
and with cuts 1--6 applied. The lines with arrows 
indicate the region selected by the  muon-electron vertex distance cut.}
\label{fig:dvse}
\end{figure}

Cut~7 utilizes the muon-electron vertex distance, the measured 
separation between the reconstructed muon and electron vertices. This
cut requires that the decay-electron is correctly associated
with the primary muon and is applied to the CCQE (2-subevent) sample only.  
This eliminates many CC1$\pip$ events where the
second subevent is a decay-positron from the $\pip$ decay chain and not 
the electron from the decay of the primary muon.
The distributions of the muon-electron vertex distance 
for the major  Monte Carlo channels and for data
are shown in Figure~\ref{fig:dvse}, after the application
of cuts 1--6 for events with 2 subevents.

As shown in Table~\ref{tab:cut}, the efficiency for finding CCQE events with
a reconstructed vertex radius, $r<550$~cm, is $\distCCQEeff$\%.  An $r<550$~cm volume
is used for normalization in the cross section calculations to correctly account 
for events that pass all cuts but have a true vertex with $r>500$~cm. Normalizing to 
events with true vertex of  $r<500$~cm yields an efficiency of 35\%.

\subsection{CC1$\pip$ background measurement}\label{sec:ccpifit}

\begin{figure}
\includegraphics[width=\columnwidth]{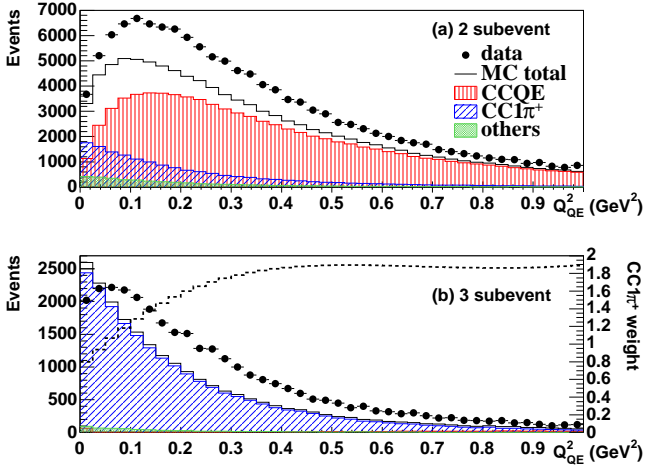}
\caption{
(color online)  
Distribution of events in $Q^2_{QE}$ for the (a) 2 and (b) 3 subevent samples
{\em before} the application of the CC1$\pip$ background correction.
Data and MC samples are shown along with the individual MC contributions from 
CCQE, CC1$\pip$, and other channels.  In (b), the dashed line shows the CC1$\pip$
reweighting function (with the y-axis scale on the right) as determined from 
the background fit procedure.}
\label{fig:fit_before}
\end{figure}

After the selection of the CCQE and CC1$\pip$ candidate events (2 and
3 subevent samples, respectively), the CC1$\pip$ background to the
CCQE signal is measured by adjusting the
weights of the simulated CC1$\pip$ events to achieve data-MC agreement
in the $Q^2_{QE}$ distribution of the 3 subevent sample.  The same
weighting, applied to all simulated CC1$\pip$ events, then provides an estimate
of the CC1$\pip$ background to the CCQE signal.
Figure~\ref{fig:fit_before} shows the data and MC $Q^2_{QE}$ distributions
for the two samples before the reweighting of CC1$\pip$ MC
events.  The 3-subevent sample is predicted to be 90\% CC1$\pip$
and shows a large data-MC disagreement in both shape and
normalization. The kinematic distribution of muons in 
CC1$\pip$ events is similar in both the 2-and 3-subevent samples
as can be observed in Fig.~\ref{fig:fit_before}.
This occurs because the majority of CC1$\pip$ events
that are background in the 2-subevent sample are due to muon-capture
or pion absorption and the reconstruction of the primary event 
is, to a good approximation, independent of this.  In addition, 
the $\mu/e$ log-likelihood ratio cut 
(Tab.~\ref{tab:cut} and Fig.~\ref{fig:lemu}) is applied for both
the 2- and 3-subevent samples, further ensuring that the CC1$\pip$ events 
are the same in both samples.

The CC1$\pip$ reweighting function (Fig.~\ref{fig:fit_before}b)
is a 4$^\mathrm{th}$-order polynomial in $Q^2_{QE}$ and is determined
from the ratio of data to MC in this sample.  The 2-subevent sample
shows good shape agreement between data and MC.  This is because the
event model for CCQE was already adjusted to match data in a previous
analysis~\cite{MB_CCQE} that considered only the shape of the
$Q^2_{QE}$ distribution.  That analysis did not consider the overall
normalization of events.

In practice, this determination of the CC1$\pip$ reweighting is done
iteratively as there is some CCQE background in the 3 subevent sample.
An overall normalization factor is calculated for the CCQE sample to
achieve data-MC agreement in the 2 subevent sample after subtraction
of the CC1$\pip$ background. This is then applied to determine the CCQE
background in the 3 subevent sample.  The background from other channels
is determined from the simulation and subtracted. This process converges 
after two iterations.

This method determines a correction to the CC1$\pip$ rate (as a function 
$Q^2_{QE}$) using data from the 3-subevent sample rather than relying strictly
on simulation.  This reweighting is then applied to all simulated CC1$\pip$ events, in 
particular those that are contained in the 2-subevent sample and
form most of the background for the CCQE measurement.  The error on 
$M_A^{1\pi}$ within the resonant background model is then set to zero and 
the resulting error on  the CC1$\pip$ background to the CCQE signal from 
CC1$\pip$ production is determined by the coherent $\pi$-production errors 
and the $\pip$ absorption uncertainty. The statistical errors in this 
procedure are negligible. Most CC1$\pip$ events that end up 
in the 2-subevent (CCQE) sample
are due to intranuclear $\pip$ absorption.  This process is modeled in the 
event simulation as explained in Sec.~\ref{sec:nufsi} and is assigned a 
25\% uncertainty. The coherent $\pi$-production process is modeled as
described in Sec.~\ref{sec:nucohpi} and is assigned a 100\% uncertainty.

\begin{figure}
\includegraphics[width=\columnwidth]{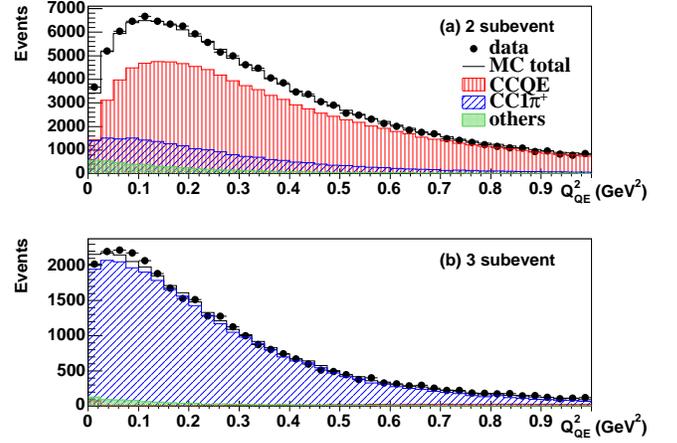}
\caption{
(color online)  
Distribution of events in $Q^2_{QE}$ for the (a) 2 and (b) 3 subevent samples
{\em after} the application of the data-based CC1$\pip$ background constraint and 
the new CCQE model parameters $\MAeff$ and $\ka$ as determined from the 
CCQE fit procedure described in the text.}
\label{fig:fit_after}
\end{figure}

With the measured CC1$\pip$ background incorporated, a shape-only fit
to the 2-subevent (CCQE) sample is performed to extract values for the
CCQE model parameters, $\MAeff$ and $\ka$.  This exercise is required
to have a consistent description of the MiniBooNE data within the
simulation after adjustment of the background.  This procedure has no
effect on the CCQE cross section results reported here other than very
small corrections to the antineutrino background subtraction which
uses these parameters.  In this fit, all systematic errors and
correlations are considered. The CCQE simulated sample is normalized
to have the same number of events as data which is the same
normalization as determined in the CC1$\pip$ background determination.
The $Q^2_{QE}$ distributions of data from the 2 and 3 subevent samples
is shown together with the MC calculation in
Figure~\ref{fig:fit_after}.  The MC calculations include all the
adjustments described in this section and agreement with data is good
in both samples.

\begin{figure}
\includegraphics[width=\columnwidth]{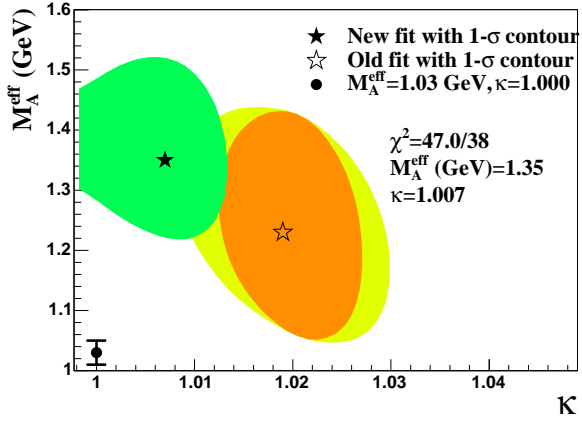}
\caption{\label{fig:contour}(Color online).  
The $1\si$ contour plot for the $\MAeff-\ka$ fit. 
The filled star shows the best fit point and $1\si$ contour 
extracted from this work. 
The open star indicates the best fit point and 
$1\si$ contour from the previous work~\cite{MB_CCQE}.
Two error ellipses are shown for the previous work, the larger reflects
the total uncertainty ultimately included in Ref.~\cite{MB_CCQE}.
The circle indicates the world-average value for $M_A$~\cite{past-ma}.} 
\end{figure}

This shape-only fit to the 2-subevent sample yields the adjusted CCQE model parameters, 
$\MAeff$ and $\ka$, 

\beq
\MAeff&=& \NEWMAcon \pm \NEWMAerr~\uGeVct~;\no\\
\ka      &=& \NEWKAcon \pm \NEWKAerr        ~;\no\\
\ch^2/\mathrm{dof}&=& \CHcon /38             ~.\no
\eeq
Figure~\ref{fig:contour} shows the $1\si$ contour regions of this fit together
with the results from the original MiniBooNE analysis~\cite{MB_CCQE}.
The new fit yields different results for both $\MAeff$ and $\ka$ because of the 
improved CC1$\pip$ background estimation method used in this analysis.  Note that the current 
result is consistent to within $1\sigma$ with $\ka=1$, unlike the previous MiniBooNE 
result. This fit actually provides no lower bound on $\ka$ as the $1\sigma$ contour
is not closed for $\ka<1$. The value for $\ka$ is quite sensitive 
to the CC1$\pip$ background at the lowest $Q^2_{QE}$ and the background in that
region has decreased in this analysis.
The increase in the CC1$\pip$ background at larger $Q^2_{QE}$ values has resulted
in a larger value for the extracted $\MAeff$.
The previous and current parameter contours are consistent at the
$1\si$ level.  Neither this nor the prior analysis result is
consistent with the world-average $M_A$ of $1.03\pm0.02$~GeV~\cite{past-ma}, 
as can be seen in Figure~\ref{fig:contour}. The $\ch^2/\mathrm{dof}$ 
assuming $M_A=1.03$~GeV, $\ka=1$ is 67.5/40 corresponding to a $\ch^2$ probability
of $\approx$ 0.5\%.

\begin{figure}
\includegraphics[width=\columnwidth]{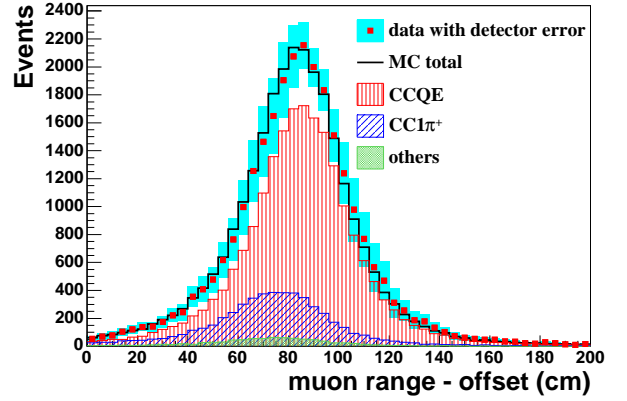}
\caption{\label{fig:tmuscale}(Color online).  
Distributions, for both data and simulation, of the measured muon range for muons 
with $400<T_\mu<500$~MeV.  The subtracted offset, ($500 \times T\mu(\mathrm{GeV})-100$)~cm, 
corresponds to the solid  diagonal line in Fig.~\ref{fig:dvse}.}
\end{figure}

The reconstructed four-momentum transfer, $Q^2_{QE}$, depends upon 
the muon energy as can be seen in Eq.~\ref{eq:recQsq}.
The reconstructed muon energy calibration has been checked by comparing 
the measured range of muons determined from the muon-electron vertex distance 
(Fig.~\ref{fig:dvse})
with the expected muon range determined from the energy provided by the 
reconstruction algorithm, which does not use this vertex distance. 
As an example, a comparison of the measured and expected muon ranges for 
muons of $400<T_\mu<500$~MeV is shown in Fig.~\ref{fig:tmuscale} for
both data and simulation.
The agreement is good for all muon energies
and verifies the energy calibration to 2\%, well within the errors calculated by the simulation.
This also shows that any light produced by hadronic particles in the neutrino
interaction (for both CCQE and background channels) is adequately simulated and considered in the reconstruction.  

\begin{figure}
\includegraphics[width=\columnwidth]{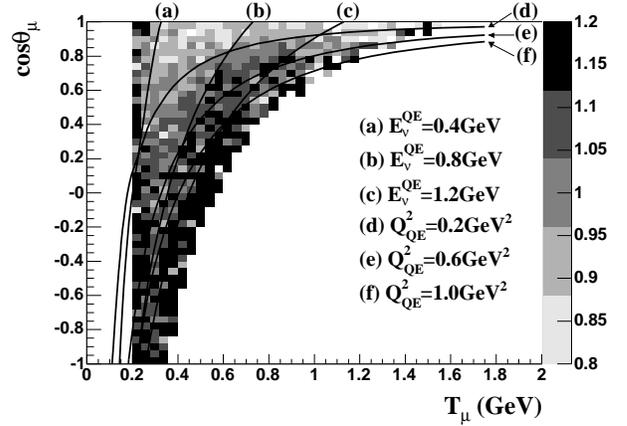}
\caption{\label{fig:2d_before}  
Ratio of $\numu$ CCQE data to simulation as a function of 
reconstructed muon kinetic energy and angle. The simulation 
includes adjustment of the CC1$\pip$ backgrounds based on MiniBooNE data
but is that prior to any CCQE model tuning (it assumes $M_A=1.03$ GeV, $\kappa=1$).
The prediction has been normalized to the data. If the simulation 
modeled the data perfectly, this 
ratio would be unity. Contours of constant $E_{\nu}$ and $Q^2$ are overlaid, 
and only bins with $>20$ events in the data are plotted.
}
\end{figure}

\begin{figure}
\includegraphics[width=\columnwidth]{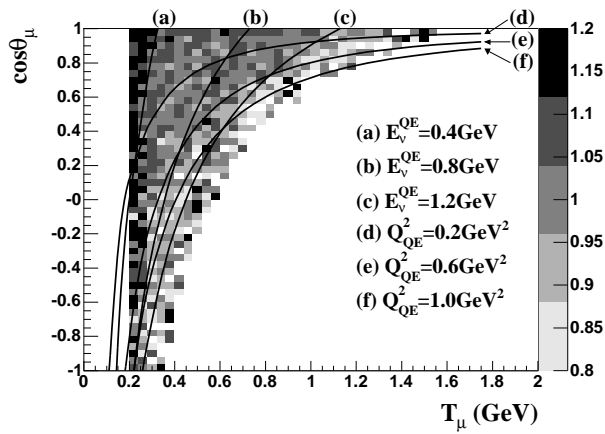}
\caption{\label{fig:2d_after}  
Ratio of $\numu$ CCQE data to simulation as a function of 
reconstructed muon kinetic energy and angle. The simulation includes
the adjusted CC1$\pip$ background prediction and the new CCQE model 
parameters ($\MAeff=1.35, \kappa=1.007$) measured from MiniBooNE data . 
Compare to Figure~\ref{fig:2d_before}.
}
\end{figure}

A final and more complete check that the simulation correctly
models the data can be made by examining the 2-dimensional muon
kinetic energy and angle ($T_\mu,\cos\th_\mu$) distributions.  While
Figure~\ref{fig:fit_after} shows that data is well-described in
$Q^2_{QE}$, the adjusted model may
not be adequate when applied to the ($T_\mu,\cos\th_\mu$) distribution of
events. This could occur if an adjustment in $Q^2_{QE}$ is hiding an
incorrect neutrino energy distribution.  The ratio of data to
simulation in ($T_\mu,\cos\th_\mu$) after correction of the CC1$\pip$
background and before the new CCQE model parameters are applied is shown in
Figure~\ref{fig:2d_before}. The ratio after all corrections is shown
in Figure~\ref{fig:2d_after} and is much closer to unity throughout
the muon phase space. As can be seen in Figure~\ref{fig:2d_before},
the regions of constant ratio mainly follow lines of constant $Q^2_{QE}$ 
and not $E_{\nu}^{QE}$. Also, almost no structure remains in 
Figure~\ref{fig:2d_after}.  The exception is band of $\approx 20\%$ disagreement
at $E_{\nu}^{QE}\approx0.4$~GeV where the error on the neutrino flux is 
of that order (Fig.~\ref{fig:nuflux}b.)
This data-simulation agreement provides additional support for our
procedure of adjusting only the $Q^2_{QE}$ behavior of the model and not 
the energy distribution of the neutrino flux.

\subsection{Extraction of the cross sections}\label{sec:xsformula}
With the CC1$\pip$ interaction background prediction determined from 
experimental data (Section~\ref{sec:ccpifit}) and the remaining channels 
predicted from the interaction model (Section~\ref{sec:nuan}), the cross
section for the CCQE interaction can be extracted. 
A total of $\evki$ events pass the
CCQE selection (Section~\ref{sec:evnt}) resulting from $\MBosc1POT$~protons 
on target (POT) collected between August 2002 and December 2005.
The efficiency for CCQE events passing these cuts is calculated to be
$\distCCQEeff$\% for CCQE events with true vertices within a 550~cm radius from
the center of the detector tank.  The sample is estimated to contain 23.0\% 
background events.  These numbers, together with a breakdown of 
predicted backgrounds, are summarized in Table~\ref{tab:CCQEevsum}. 

\begin{table}
\begin{tabular}{lrr}
\hline
\hline
integrated protons on target   & \multicolumn{2}{r}{$\MBosc1POT$}       \\
energy-integrated $\numu$ flux & \multicolumn{2}{r}{$\MBnumuflxint$ $\numu$/cm$^2$} \\
CCQE candidate events          & \multicolumn{2}{r}{$\evki$}           \\
CCQE efficiency ($R<550$~cm)   & \multicolumn{2}{r}{$\distCCQEeff$\%}  \\
\hline
background channel             &  events   & fraction    \\
\hline
NCE                            &  $\eNCEL$ & $\fNCELs$\% \\
CC1$\pip$                      &  $\eCCPP$ & $\fCCPPs$\% \\
CC1$\piz$                      &  $\eCCP0$ & $\fCCP0s$\% \\
NC1$\pi^\pm$                   &  $\eNCPP$ & $\fNCPPs$\% \\
NC1$\piz$                      &  $\eNCP0$ & $\fNCP0s$\% \\
other $\numu$                  &  $\eOTHR$ & $\fOTHRs$\% \\
all non-$\numu$                &  $\eANTI$ & $\fANTIs$\% \\
\hline
total background               &  $\eTBCK$ & $\fTBCKs$\% \\  
\hline
\hline
\end{tabular}
\caption{Summary of the final CCQE event sample including a breakdown of the 
estimated backgrounds from individual channels. The fraction is relative to 
the total measured sample. The channel nomenclature is defined in 
Table~\ref{tab:pichart}.}
\label{tab:CCQEevsum}
\end{table}

The background is dominated by CC1$\pip$ interactions which are estimated
to be 18.4\% of the CCQE candidate sample.  Their predicted distribution
in $Q^2_{QE}$ is shown in Figure~\ref{fig:fit_after}(a). As can be seen, this 
background is a substantial fraction of the sample in the lowest $Q^2_{QE}$ 
region. The majority (52\%) of the CC1$\pip$ background is predicted to be 
events in which the $\pip$ is absorbed in the initial target nucleus. These 
are defined as ``CCQE-like'' in that they contain a muon and no pions in the 
final state. The remaining CC1$\pip$ background consists of CC1$\pip$ events 
where the $\pip$ is absorbed outside of the target nucleus (33\%),  
is not identified due to a missed muon decay (11\%), or 
undergoes charge exchange in the nucleus or detector medium (4\%),  
These last three classes of CC1$\pip$ 
backgrounds are not considered CCQE-like. All CC1$\pip$ background events,
including those that are CCQE-like, are subtracted from the data to obtain 
the final CCQE cross section results. However, to facilitate examination of 
the model used for these processes, the effective cross section for CCQE-like 
background events is separately reported in the Appendix.

In this analysis, the small contamination of $\numub$, $\nue$, and $\nueb$ 
interactions are treated as background and are subtracted from the data 
based on their MC prediction (see Tab.~\ref{tab:CCQEevsum}).
The majority of these are $\numub$ CCQE interactions. The same 
$\MAeff$ and $\ka$ as measured in the $\nu_\mu$ CCQE sample are used 
to predict non-$\numu$ CCQE events. These parameters have been shown to 
adequately reproduce the MiniBooNE CCQE data collected in antineutrino 
mode~\cite{JoeG}.

To extract differential cross sections in muon kinematic variables, the 
reconstructed kinematics are corrected for detector-specific effects.
A correction procedure is implemented using an ``unfolding'' process based 
on the detector simulation. We employ an ``iterative Bayesian'' 
method~\cite{DAgostini} to avoid the problem of amplification of statistical 
fluctuations common in the ``inverse response matrix'' method~\cite{stattext}.
A disadvantage to the iterative Bayesian method is that the result depends 
on the initial CCQE model assumptions (the ``prior'' probability).  However, 
this problem is addressed by an iterative method that uses the extracted 
signal distribution to correct the predicted distributions and repeating this 
procedure.  In practice, the simulation was already tuned to reproduce the 
data based on previous work~\cite{MB_CCQE}, and the result from the first 
iteration shows satisfactory convergence.  The systematic uncertainty in 
this procedure is determined from the difference between the initial (0th) 
and final (1st) iterations of the algorithm and by examining the dependence 
of the final values on the initial model assumptions.

The various correction and normalization factors can then be brought together
in a single expression used to extract the CCQE cross section for the $i$th bin of a 
particular kinematic variable,  
\begin{equation}
\si_i=\fr{\sum_{j}U_{ij}(d_j-b_j)}{\ep_i\cdot T\cdot \Ph},
\label{eq:xseq}
\end{equation}
where the index $j$ labels the reconstructed bin and $i$ labels the unfolded 
(estimate of the true) bin. In this equation, $U_{ij}$ is the unfolding 
matrix, $\ep_i$ is the efficiency, $T$ is the number of neutrons in the 
fiducial volume, and $\Ph$ is the neutrino flux.  This expression is used to 
obtain the double and single differential cross sections,  
$\frac{d^2\si}{dT_\mu d\cos\th_\mu}$, and $\frac{d\si}{dQ^2_{QE}}$, 
respectively, after the multiplication of the appropriate
bin width factors. Note that the choice of normalization yields cross sections
``per neutron''. Here, the flux, $\Ph$, is a single number ($\MBnumuflxint$~$\numu$/cm$^2$)
and is determined by integrating the BNB flux over $0<E_\nu<3$~GeV.  Therefore, these differential
cross sections are ``flux-integrated''.

An additional quantity, the flux-unfolded CCQE cross section as a function of
neutrino energy, $\si[E_\nu^{QE,RFG}]$, is extracted from this same 
expression with the replacement of the total flux, $\Ph$, with the flux in 
a particular neutrino energy bin, $\Ph_i$. 
The unfolding procedure is used to correct the data from bins of reconstructed 
neutrino energy, $E_\nu^{QE}$, (using Eq.~\ref{eq:recEnu}) to an estimate of the true 
neutrino energy, $E_\nu^{QE,RFG}$.  It is important to note that, unlike the differential 
cross sections, $\frac{d^2\si}{dT_\mu d\cos\th_\mu}$ and 
$\frac{d\si}{dQ^2_{QE}}$, the calculation of this cross section relies on the 
interaction model to connect $E_\nu^{QE}$ to $E_\nu^{QE,RFG}$.  The superscript
``$RFG$'' indicates the interaction model assumed in the unfolding process~\cite{SmithMoniz}. 
This procedure introduces a model dependence into this cross section, however, 
this method is consistent with that commonly used by experiments reporting a CCQE 
cross section as a function of neutrino energy.  This model-dependence should 
be considered when comparing measurements of this quantity from different experiments.

\subsection{Error analysis}\label{sec:errmat}
The errors on the measured cross sections result from uncertainties in the neutrino flux, 
background estimates, detector response, and unfolding procedure. To propagate
these error sources, a ``multisim'' method~\cite{Teppei} is used to calculate the errors 
on the final quantities by varying parameters in separate simulations. This 
method produces an error matrix, $V_{ij}$, for each reported distribution that
can then be used to extract the error on each bin ($\si_{i}=\sqrt{V_{ii}}$) 
and the correlations between quantities of different bins. 

The error matrix is calculated by generating a large 
number of simulated data sets with different parameter excursions, based on 
the estimated 1$\sigma$ uncertainties in those parameters and the 
correlations between them.  The error matrix for a particular distribution is 
then calculated from these $M$ data sets,  
\begin{equation}
V_{ij} = \fr{1}{M}\sum_{s=1}^{M}(Q_i^{s}-\hat{Q}_i)(Q_j^{s}-\hat{Q}_j).
\label{eq:errmat}
\end{equation}
Here, $Q_i^{s}$ is the quantity of interest in the $i^{th}$ bin from the 
$s^{th}$ simulation data set and $\hat{Q}_i$ is the ``best'' estimate of the 
parameters. The quantities of interest $Q_i$ could be, for instance, the 
number of events in  each bin of $Q^2_{QE}$ or the calculated cross 
section in each kinematic bin  (Eq.~\ref{eq:xseq}).

In practice, the errors are classified into four major contributions: 
the neutrino flux, background cross sections, detector modeling, and 
unfolding procedure.  The parameters within each of these groups are varied
independently so the resultant error matrices from the individual groups 
can be added to form the total error matrix. For the neutrino 
flux and background cross section uncertainties, a reweighting method is 
employed which removes the difficulty of requiring hundreds of simulations 
with adequate statistics.  In this method, each neutrino interaction event is 
given a new weight calculated with a particular parameter excursion. This is 
performed considering correlations between parameters and allows each 
generated event to be re-used many times saving significant CPU time.   
The nature of the detector uncertainties does not allow for this method of
error evaluation as parameter uncertainties can only be applied as each 
particle or optical photon propagates through the detector. Approximately 
100 different simulated data sets are generated with the detector parameters 
varied according to the estimated $1\sigma$ errors including correlations. 
Eq.~\ref{eq:errmat} is then used to calculate the detector error matrix. The 
error on the unfolding procedure is calculated from the difference in 
final results when using different input model 
assumptions (Section~\ref{sec:xsformula}). The statistical error on 
data is not added explicitly but is included via the statistical fluctuations 
of the simulated data sets (which have the same number of events as the data).

The final uncertainties are reported in the following sections. The 
breakdown among the various contributions are summarized and discussed in 
Section~\ref{sec:syserr}.  For simplicity, the full error matrices are not 
reported for all distributions. Instead, the errors are separated into a total
normalization error, which is an error on the overall scale of the cross 
section, and a ``shape error'' which contains the uncertainty that does 
not factor out into a scale error. This allows for a distribution of data to 
be used (e.g. in a model fit) with an overall scale error for uncertainties
that are completely correlated between bins, together with the remaining 
bin-dependent shape error.

\section{Results and discussion}\label{sec:xsec}

\subsection{CCQE flux-integrated double differential cross section}\label{sec:ddsigma}
\begin{figure}
\includegraphics[width=\columnwidth]{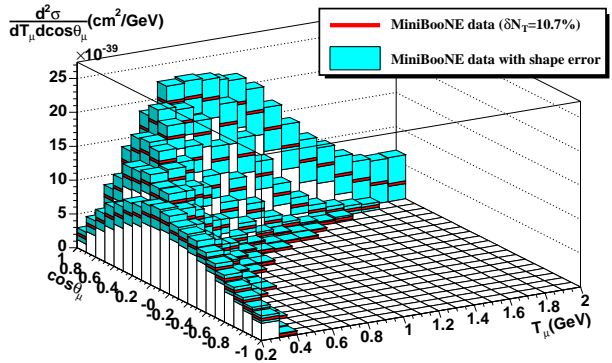}
\caption{\label{fig:ddsigma}(Color online).  
Flux-integrated double differential cross section per
target neutron for
the $\nu_\mu$ CCQE process.  The dark bars indicate the measured values and
the surrounding lighter bands show the shape error.  The overall normalization 
(scale) error is $\ttXSerr$\%. Numerical values are provided in 
Table~\ref{tab:ddsigma} in the Appendix.}
\end{figure}

The flux-integrated, double differential cross section per neutron, 
$\frac{d^2\si}{dT_\mu d\cos\th_\mu}$, for the $\nu_\mu$ CCQE process is 
extracted as described in Section~\ref{sec:xsformula} and is shown in 
Figure~\ref{fig:ddsigma} for the kinematic range, $-1<\cos\th_\mu<+1$, 
$0.2<T_\mu(\uGeV)<2.0$.  The errors, for $T_\mu$ outside of this range, 
are too large to allow a measurement.  Also, bins with low event population near or outside of the 
kinematic edge of the distribution (corresponding to large $E_\nu$) do not allow for 
a measurement and are shown as zero in the plot. The numerical values for this
double differential cross section are provided in Table~\ref{tab:ddsigma} in 
the Appendix.

The flux-integrated CCQE total cross section, obtained by integrating the double 
differential cross section (over $-1<\cos\th_\mu<+1$, $0<T_\mu(\uGeV)<\infty$), 
is measured to be $\absXScon\times 10^{-39}$~$\ucmt$. The total normalization error on 
this measurement is $\ttXSerr$\%.

The kinematic quantities, $T_\mu$ and $\cos\th_\mu$, have been corrected for 
detector resolution effects only (Section~\ref{sec:xsformula}).  Thus, this result 
is the most model-independent measurement of this process possible with the 
MiniBooNE detector. No requirements on the nucleonic final state are used to define
this process. The neutrino flux is an absolute prediction~\cite{beam} and has not been
adjusted based on measured processes in the MiniBooNE detector.

\subsection{Flux-integrated single differential cross section}\label{sec:dsigma}
\begin{figure}
\includegraphics[width=\columnwidth]{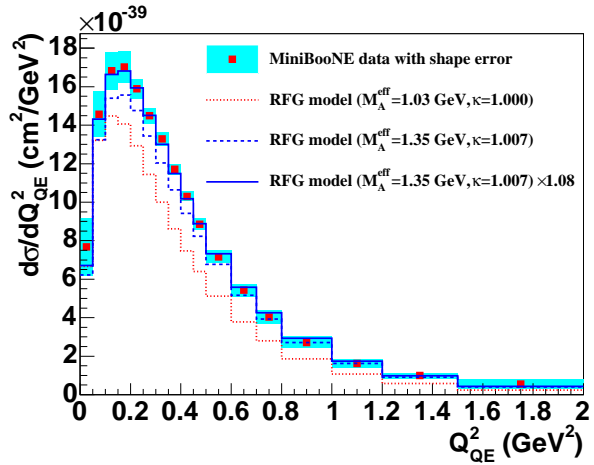}
\caption{\label{fig:dsigma}(Color online).  
Flux-integrated single differential cross section per target neutron 
for the $\nu_\mu$ CCQE process. The measured values are shown as points 
with the shape error as shaded bars. Calculations from the \nuance\ RFG
model with different assumptions for the model parameters are shown as 
histograms. Numerical values are provided in Table~\ref{tab:dsigma} in 
the Appendix.}
\end{figure}

The flux-integrated, single differential cross section per neutron, 
$\frac{d\si}{dQ^2_{QE}}$, has also been measured and is shown in 
Figure.~\ref{fig:dsigma}. The quantity $Q^2_{QE}$ is defined in Eq.~\ref{eq:recQsq} and depends
only on the (unfolded) quantities $T_\mu$ and $\cos\th_\mu$.  
It should be noted that the efficiency for events with  
$T_\mu<200$~MeV is not zero because of difference between reconstructed
and unfolded $T_\mu$.  The calculation of efficiency for these (low-$Q^2_{QE}$)
events depends only on the model of the detector response, not on an interaction
model and the associated uncertainty is propagated to the reported results.

In addition to the experimental result, Figure~\ref{fig:dsigma} also shows 
the prediction for the CCQE process from the \nuance\ simulation with three
different sets of parameters in the underlying RFG model.  The 
predictions are absolutely normalized and have been integrated over the
MiniBooNE flux.
The RFG model is plotted assuming both the world-averaged CCQE parameters 
($M_A=\NUAMAcon~\uGeV$, $\ka=\NUAKAcon$)~\cite{past-ma} and the CCQE 
parameters extracted from this analysis ($M_A=\NEWMAcon~\uGeV$, 
$\ka=\NEWKAcon$) in a shape-only fit. The model
using the world-averaged CCQE parameters underpredicts the 
measured differential cross section values by $20-30\%$, while the model 
using the CCQE parameters extracted  from this shape analysis are within 
$\approx 8$\% of the data, consistent within 
the normalization error ($\approx 10$\%). To further illustrate this, the 
model calculation with the CCQE parameters from this analysis scaled by 1.08 is also 
plotted and shown to be in good agreement with the data.

\subsection{Flux-unfolded CCQE cross section as a function of neutrino energy}\label{sec:sigma}

\begin{figure}
\includegraphics[width=\columnwidth]{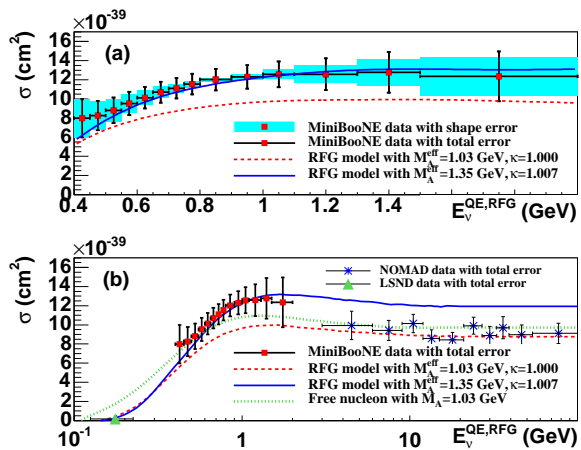}
\caption{\label{fig:sigma}(Color online).  
Flux-unfolded MiniBooNE $\nu_\mu$ CCQE cross section per neutron as a 
function of neutrino energy. In (a), shape errors are shown as shaded boxes
along with the total errors as bars. 
In (b), a larger energy range is shown along with results from the 
LSND~\cite{LSNDxs} and NOMAD~\cite{NOMAD} experiments.
Also shown are predictions from the \nuance\ simulation for an RFG
model with two different parameter variations and for scattering from free nucleons
with the world-average $M_A$ value. Numerical values are provided in Table~\ref{tab:sigma} 
in the Appendix.}
\end{figure}

The flux-unfolded CCQE cross section per neutron, $\si[E_\nu^{QE,RFG}]$, 
as a function of the true neutrino energy, $E_\nu^{QE,RFG}$, is shown in 
Figure~\ref{fig:sigma}. These numerical values are tabulated in
Table~\ref{tab:sigma} in the Appendix.
The quantity $E_\nu^{QE,RFG}$ is a (model-dependent) estimate of the 
neutrino energy obtained after correcting for both detector and nuclear model
resolution effects.  These results depend on the details of the nuclear model
used for the calculation. The dependence is only weak in the peak of the flux
distribution but becomes strong for $E_\nu<0.5$~GeV and $E_\nu>1.2$~GeV,
i.e., in the ``tails'' of the flux distribution.  

In Figure~\ref{fig:sigma}, the data are compared with the \nuance\
implementation of the RFG model with the world average parameter values, 
($\MAeff=\NUAMAcon~\uGeV$, $\ka=\NUAKAcon$) and with the parameters extracted 
from this work ($\MAeff=\NEWMAcon~\uGeV$, $\ka=\NEWKAcon$). These are absolute
predictions from the model (not scaled or renormalized).
At the average energy of the MiniBooNE flux ($\approx 800$~MeV), the extracted cross section 
is $\approx 30$\% larger than the RFG 
model prediction with world average parameter values.  The RFG model,
with parameter values extracted from the {\em shape-only} fit to this data  
better reproduces the data over the entire measured energy range. 

Figure~\ref{fig:sigma}(b) shows these CCQE results together with those from 
the LSND~\cite{LSNDxs} and NOMAD~\cite{NOMAD} experiments. It is interesting 
to note that the NOMAD results are better described with the world-average 
$\MAeff$ and $\ka$ values.  Also shown for comparison in Fig.~\ref{fig:sigma}(b) 
is the predicted cross section assuming the CCQE interaction occurs on 
free nucleons with the world-average $M_A$ value.  The cross sections reported
here exceed the free nucleon value for $E_\nu$ above 0.7~GeV.  

\subsection{Error Summary}\label{sec:syserr} 

\begin{table}
\begin{tabular}{lr}
\hline
\hline
source                   &normalization error (\%)\\
\hline
neutrino flux prediction           &$\flXSerr$\\
background cross sections          &$\xtXSerr$\\
detector model                     &$\deXSerr$\\
kinematic unfolding procedure      &$\itXSerr$\\
statistics                         &$\stXSerr$ \\
\hline
total                              &$\ttXSerr$\\
\hline
\hline
\end{tabular}
\caption{Contribution to the total normalization uncertainty from each of the 
various systematic error categories.}
\label{tab:syserr}
\end{table}

\begin{figure}
\includegraphics[width=\columnwidth]{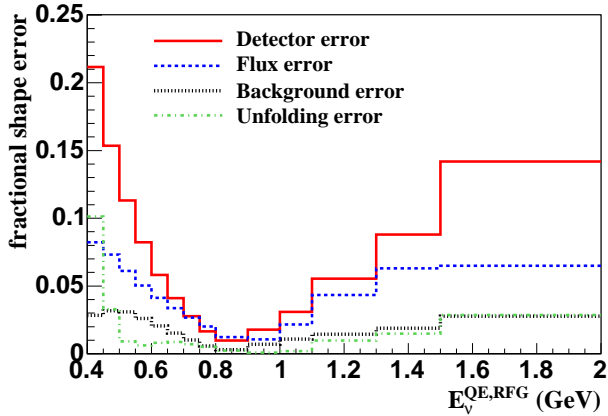}
\caption{\label{fig:sigma_sys} 
Fractional shape error on the MiniBooNE $\nu_\mu$ CCQE flux-unfolded 
cross section separated into major components. The overall normalization
error of 10.7\% is not shown.}
\end{figure}

As described in Section~\ref{sec:errmat}, (correlated) systematic and 
statistical errors are propagated to the final results.  These errors are 
separated into normalization and shape uncertainties.  The contributions from 
each error source on the total normalization uncertainty 
are summarized in Table~\ref{tab:syserr}. As is evident, the neutrino flux 
uncertainty dominates the overall normalization error on the extracted 
CCQE cross sections. However, the uncertainty on the flux prediction is a 
smaller contribution to the shape error on the cross sections.  This can be 
seen in Figure~\ref{fig:sigma_sys} which shows the contribution from the 
four major sources to the shape error on the total (flux-unfolded) 
cross section.
  
The detector model uncertainty dominates the shape error, especially at low 
and high energies. This is because errors in the detector response (mainly via 
uncertainties in visible photon processes) will result in errors on the 
reconstructed energy. These errors grow in the tails of the neutrino flux 
distribution due to feed-down from events in the flux peak.  This type of 
measurement usually has large errors due to non-negligible uncertainties
in the CC1$\pip$ background predictions.  In this measurement, that error 
is reduced through direct measurement of the CC1$\pip$ background.  However, 
this error is not completely eliminated due to the residual uncertainty 
on the rate of intranuclear pion absorption that is included. This 
uncertainty is not as important for the measurement of the CCQE cross 
section measurement as a function of energy but is a large contribution
to the error at low $Q^2_{QE}$ in the differential distributions. 

The unfolding error is small in the 
region of the flux peak but grows in the high- and low-energy region because 
of the uncertainty in the feed-down from other energy bins, similar to that
described for the detector model. 

\section{Conclusions}\label{sec:conclu}
In this work, we report measurements of absolute cross sections for
the CCQE interaction using high-statistics samples of $\nu_\mu$ interactions 
on carbon. These include the first measurement of the double differential cross section, 
$\frac{d^2\si}{dT_\mu d\cos\th_\mu}$, measurement of the single differential
cross section, $\frac{d\si}{dQ^2_{QE}}$, and the flux-unfolded CCQE
cross section, $\si[E_\nu^{QE,RFG}]$. The double differential cross section
contains the most complete and model-independent information that is
available from MiniBooNE on the CCQE process. It is the main result from 
this work and should be used 
as the preferred choice for comparison to theoretical models of CCQE 
interactions on nuclear targets.
 
The reported cross section is significantly larger ($\approx 30$\% at the
flux average energy) than what is commonly 
assumed for this process assuming a relativistic Fermi Gas model (RFG) and the 
world-average value for the axial mass, $M_A=1.03$~GeV~\cite{past-ma}. In addition, 
the $Q^2_{QE}$ distribution of this data shows a significant excess of events 
over this expectation at higher $Q^2_{QE}$ even if the data is normalized to 
the prediction over all $Q^2_{QE}$. This leads to an extracted axial mass from 
a ``shape-only'' fit of the $Q^2_{QE}$ distribution of $\MAeff=1.35\pm0.17$~GeV,
significantly higher than the historical world-average value.

These two observations, unexpectedly large values for the extracted cross section 
and $\MAeff$, are experimentally separate. However, within the model 
prediction, a larger value for $M_A$ implies a larger cross section because 
the CCQE cross section increases approximately linearly with $M_A$.  The predicted 
CCQE cross section with this higher value of $M_A$ agrees with the measurement
within the normalization error of the experiment ($\approx 10\%$).  While this
may be simply a coincidence, it is important to note. 

In recent years, there has been significant effort to improve the theoretical 
description of the CCQE interaction on nuclear targets~\cite{new-model,Martini};
however, there seems to be no simple explanation for both the higher cross section
and the harder $Q^2_{QE}$ distribution of events (resulting in a larger 
$\MAeff$) as evidenced by the MiniBooNE CCQE data. Nuclear effects can have 
some impact on the measured $\MAeff$, but it is not obvious that they are 
large enough. Also, it is expected that such effects should reduce the cross section, not 
increase it.  This can be be seen in Fig.~\ref{fig:sigma} where the cross section 
for the CCQE interaction on free nucleons is compared to that from bound nucleons
(in an RFG model). Note the reduction in cross section from free to bound nucleons.
It is interesting that the MiniBooNE measurement is also larger than this free
nucleon value (at least at higher energies).  This may indicate a significant 
contribution from neglected mechanisms for CCQE-like scattering from a nucleus such
as multi-nucleon processes (for example, Ref.~\cite{Martini}). This may 
explain both the higher cross section and the harder $Q^2$ spectrum, but has not 
yet been explicitly tested. It may also be relevant for the 
difference between these results and those of NOMAD (or other experiments) where
the observation of recoil nucleons enter the definition of a CCQE event.
An important test for such models will be their ability 
to accurately reproduce the MiniBooNE double differential cross sections at 
least as well as the RFG model assuming a higher axial mass value. 

As yet, there is no easily recognized solution to explain the difference 
between the CCQE cross sections measured in MiniBooNE at lower neutrino 
energy ($E_\nu < 2$ GeV) and the NOMAD results at higher neutrino energies
($E_\nu > 3$ GeV). Model-independent measurements of the CCQE cross section 
anticipated from SciBooNE~\cite{sciboone}, MicroBooNE~\cite{muboone}, 
and MINERvA~\cite{MINERvA} as well as the T2K~\cite{T2K} 
and NOvA~\cite{NOvA} near detectors running 
with $2<E_\nu<20$~GeV, will be important to help resolve these results.

\section{Acknowledgments}
This work was conducted with support from Fermilab, the U.S. Department of Energy,
and the National Science Foundation.

\appendix
\section{Tabulation of results} 
This appendix contains tables of numerical values corresponding to various 
plots appearing in the main body of the paper.  In addition, the effective 
cross section for the CCQE-like background to the CCQE measurement is 
tabulated.  These tables are also available via the MiniBooNE website~\cite{MB-www-data}.

\subsection{Predicted $\nu_\mu$ flux}\label{app:nuflux}
Table~\ref{tab:nuflux} lists the predicted $\nu_\mu$ flux 
(Figure~\ref{fig:nuflux}) at the MiniBooNE detector in 50~MeV-wide neutrino 
energy bins.  The flux is normalized to protons on target (POT).  The mean 
energy is 788~MeV and the integrated flux over the energy range 
($0.0<E_\nu<3.0$~GeV) is $\MBnumuflx$~$\numu$/POT/cm$^2$. For this analysis, 
the total POT collected is $\MBosc1POT$ yielding an integrated flux 
of $\MBnumuflxint$~$\numu$/cm$^2$.

\begin{table*}
\begin{tabular}{cccc|cccc|cccc}
\hline
\hline
&$E_\nu$ bin && $\nu_\mu$ flux           &&$E_\nu$ bin && $\nu_\mu$ flux           &&$E_\nu$ bin && $\nu_\mu$ flux \\
&(GeV)       && ($\numu$/POT/GeV/cm$^2$) &&(GeV)       && ($\numu$/POT/GeV/cm$^2$) &&(GeV)       && ($\numu$/POT/GeV/cm$^2$) \\
\hline
&0.00-0.05&&$4.54\times 10^{-11}$&&1.00-1.05&&$3.35\times 10^{-10}$&&2.00-2.05&&$1.92\times 10^{-11}$\\
&0.05-0.10&&$1.71\times 10^{-10}$&&1.05-1.10&&$3.12\times 10^{-10}$&&2.05-2.10&&$1.63\times 10^{-11}$\\
&0.10-0.15&&$2.22\times 10^{-10}$&&1.10-1.15&&$2.88\times 10^{-10}$&&2.10-2.15&&$1.39\times 10^{-11}$\\
&0.15-0.20&&$2.67\times 10^{-10}$&&1.15-1.20&&$2.64\times 10^{-10}$&&2.15-2.20&&$1.19\times 10^{-11}$\\
&0.20-0.25&&$3.32\times 10^{-10}$&&1.20-1.25&&$2.39\times 10^{-10}$&&2.20-2.25&&$1.03\times 10^{-11}$\\
&0.25-0.30&&$3.64\times 10^{-10}$&&1.25-1.30&&$2.14\times 10^{-10}$&&2.25-2.30&&$8.96\times 10^{-12}$\\
&0.30-0.35&&$3.89\times 10^{-10}$&&1.30-1.35&&$1.90\times 10^{-10}$&&2.30-2.35&&$7.87\times 10^{-12}$\\
&0.35-0.40&&$4.09\times 10^{-10}$&&1.35-1.40&&$1.67\times 10^{-10}$&&2.35-2.40&&$7.00\times 10^{-12}$\\
&0.40-0.45&&$4.32\times 10^{-10}$&&1.40-1.45&&$1.46\times 10^{-10}$&&2.40-2.45&&$6.30\times 10^{-12}$\\
&0.45-0.50&&$4.48\times 10^{-10}$&&1.45-1.50&&$1.26\times 10^{-10}$&&2.45-2.50&&$5.73\times 10^{-12}$\\
&0.50-0.55&&$4.56\times 10^{-10}$&&1.50-1.55&&$1.08\times 10^{-10}$&&2.50-2.55&&$5.23\times 10^{-12}$\\
&0.55-0.60&&$4.58\times 10^{-10}$&&1.55-1.60&&$9.20\times 10^{-11}$&&2.55-2.60&&$4.82\times 10^{-12}$\\
&0.60-0.65&&$4.55\times 10^{-10}$&&1.60-1.65&&$7.80\times 10^{-11}$&&2.60-2.65&&$4.55\times 10^{-12}$\\
&0.65-0.70&&$4.51\times 10^{-10}$&&1.65-1.70&&$6.57\times 10^{-11}$&&2.65-2.70&&$4.22\times 10^{-12}$\\
&0.70-0.75&&$4.43\times 10^{-10}$&&1.70-1.75&&$5.52\times 10^{-11}$&&2.70-2.75&&$3.99\times 10^{-12}$\\
&0.75-0.80&&$4.31\times 10^{-10}$&&1.75-1.80&&$4.62\times 10^{-11}$&&2.75-2.80&&$3.84\times 10^{-12}$\\
&0.80-0.85&&$4.16\times 10^{-10}$&&1.80-1.85&&$3.86\times 10^{-11}$&&2.80-2.85&&$3.63\times 10^{-12}$\\
&0.85-0.90&&$3.98\times 10^{-10}$&&1.85-1.90&&$3.23\times 10^{-11}$&&2.85-2.90&&$3.45\times 10^{-12}$\\
&0.90-0.95&&$3.79\times 10^{-10}$&&1.90-1.95&&$2.71\times 10^{-11}$&&2.90-2.95&&$3.33\times 10^{-12}$\\
&0.95-1.00&&$3.58\times 10^{-10}$&&1.95-2.00&&$2.28\times 10^{-11}$&&2.95-3.00&&$3.20\times 10^{-12}$\\
\hline
\hline
\end{tabular}
\caption{
Predicted $\nu_\mu$ flux at the MiniBooNE detector.}
\label{tab:nuflux}
\end{table*}

\subsection{CCQE flux-integrated double differential cross section}\label{app:ddsigma}
Table~\ref{tab:ddsigma} contains the flux-integrated $\nu_\mu$ CCQE double 
differential cross section values ($\frac{d^2\si}{dT_\mu d\cos\th_\mu}$) in 
bins of muon energy, $T_\mu$, and cosine of the muon scattering angle with 
respect to the incoming neutrino direction (in the lab frame), $\cos\th_\mu$.
These values correspond to the plot of Figure~\ref{fig:ddsigma}. 
The integrated value over the region ( $-1<\cos\th_\mu<+1$ and 
$0<T_\mu<\infty$) is $\absXScon\times10^{-39}~\ucmt$. The total normalization 
uncertainty is $\ttXSerr$\%. Table~\ref{tab:ddsigma_err} present an analagous 
summary of the shape error for each bin.

\begin{table*}
\begin{center}
\scriptsize\addtolength{\tabcolsep}{-0.8pt}
\begin{tabular}{c|cccccccccccccccccc}
\hline
\hline
\raisebox{-0.5ex}{$\cos\th_\mu$}\raisebox{0.5ex}{$T_\mu$(GeV)}&0.2,0.3&0.3,0.4&0.4,0.5&0.5,0.6&0.6,0.7&0.7,0.8&0.8,0.9&0.9,1.0&
                                              1.0,1.1&1.1,1.2&1.2,1.3&1.3,1.4&1.4,1.5&1.5,1.6&1.6,1.7&1.7,1.8&1.8,1.9&1.9,2.0\\
\hline
+0.9,+1.0& 190.0& 326.5& 539.2& 901.8&  1288&  1633&  1857&  1874&  1803&  1636&  1354&  1047& 794.0& 687.9& 494.3& 372.5& 278.3& 227.4\\
+0.8,+0.9& 401.9& 780.6&  1258&  1714&  2084&  2100&  2035&  1620&  1118& 783.6& 451.9& 239.4& 116.4& 73.07& 41.67& 36.55&  --- &  --- \\
+0.7,+0.8& 553.6& 981.1&  1501&  1884&  1847&  1629&  1203& 723.8& 359.8& 156.2& 66.90& 26.87& 1.527& 19.50&  --- &  --- &  --- &  --- \\
+0.6,+0.7& 681.9&  1222&  1546&  1738&  1365& 909.6& 526.7& 222.8& 81.65& 35.61& 11.36& 0.131&  --- &  --- &  --- &  --- &  --- &  --- \\
+0.5,+0.6& 765.6&  1233&  1495&  1289& 872.2& 392.3& 157.5& 49.23& 9.241& 1.229& 4.162&  --- &  --- &  --- &  --- &  --- &  --- &  --- \\
+0.4,+0.5& 871.9&  1279&  1301& 989.9& 469.1& 147.4& 45.02& 12.44& 1.012&  --- &  --- &  --- &  --- &  --- &  --- &  --- &  --- &  --- \\
+0.3,+0.4& 910.2&  1157&  1054& 628.8& 231.0& 57.95& 10.69&  --- &  --- &  --- &  --- &  --- &  --- &  --- &  --- &  --- &  --- &  --- \\
+0.2,+0.3& 992.3&  1148& 850.0& 394.4& 105.0& 16.96& 10.93&  --- &  --- &  --- &  --- &  --- &  --- &  --- &  --- &  --- &  --- &  --- \\
+0.1,+0.2&  1007& 970.2& 547.9& 201.5& 36.51& 0.844&  --- &  --- &  --- &  --- &  --- &  --- &  --- &  --- &  --- &  --- &  --- &  --- \\
 0.0,+0.1&  1003& 813.1& 404.9& 92.93& 11.63&  --- &  --- &  --- &  --- &  --- &  --- &  --- &  --- &  --- &  --- &  --- &  --- &  --- \\
-0.1, 0.0& 919.3& 686.6& 272.3& 40.63& 2.176&  --- &  --- &  --- &  --- &  --- &  --- &  --- &  --- &  --- &  --- &  --- &  --- &  --- \\
-0.2,-0.1& 891.8& 503.3& 134.7& 10.92& 0.071&  --- &  --- &  --- &  --- &  --- &  --- &  --- &  --- &  --- &  --- &  --- &  --- &  --- \\
-0.3,-0.2& 857.5& 401.6& 79.10& 1.947&  --- &  --- &  --- &  --- &  --- &  --- &  --- &  --- &  --- &  --- &  --- &  --- &  --- &  --- \\
-0.4,-0.3& 778.1& 292.1& 33.69&  --- &  --- &  --- &  --- &  --- &  --- &  --- &  --- &  --- &  --- &  --- &  --- &  --- &  --- &  --- \\
-0.5,-0.4& 692.3& 202.2& 17.42&  --- &  --- &  --- &  --- &  --- &  --- &  --- &  --- &  --- &  --- &  --- &  --- &  --- &  --- &  --- \\
-0.6,-0.5& 600.2& 135.2& 3.624&  --- &  --- &  --- &  --- &  --- &  --- &  --- &  --- &  --- &  --- &  --- &  --- &  --- &  --- &  --- \\
-0.7,-0.6& 497.6& 85.80& 0.164&  --- &  --- &  --- &  --- &  --- &  --- &  --- &  --- &  --- &  --- &  --- &  --- &  --- &  --- &  --- \\
-0.8,-0.7& 418.3& 44.84&  --- &  --- &  --- &  --- &  --- &  --- &  --- &  --- &  --- &  --- &  --- &  --- &  --- &  --- &  --- &  --- \\
-0.9,-0.8& 348.7& 25.82&  --- &  --- &  --- &  --- &  --- &  --- &  --- &  --- &  --- &  --- &  --- &  --- &  --- &  --- &  --- &  --- \\
-1.0,-0.9& 289.2& 15.18&  --- &  --- &  --- &  --- &  --- &  --- &  --- &  --- &  --- &  --- &  --- &  --- &  --- &  --- &  --- &  --- \\
\hline
\hline
\end{tabular}
\end{center}
\caption{
The MiniBooNE $\nu_\mu$ CCQE flux-integrated double differential cross section 
in units of $10^{-41}~\ucmt/\uGeV$ in 0.1~GeV bins of $T_\mu$ (columns) and 
0.1~bins of $\cos\th_\mu$ (rows).}
\label{tab:ddsigma}
\end{table*}

\begin{table*}
\begin{center}
\scriptsize\addtolength{\tabcolsep}{-0.8pt}
\begin{tabular}{c|cccccccccccccccccc}
\hline
\hline
\raisebox{-0.5ex}{$\cos\th_\mu$}\raisebox{0.5ex}{$T_\mu$(GeV)}&0.2,0.3&0.3,0.4&0.4,0.5&0.5,0.6&0.6,0.7&0.7,0.8&0.8,0.9&0.9,1.0&
                                              1.0,1.1&1.1,1.2&1.2,1.3&1.3,1.4&1.4,1.5&1.5,1.6&1.6,1.7&1.7,1.8&1.8,1.9&1.9,2.0\\
\hline
+0.9,+1.0& 684.3&  1071&  1378&  1664&  1883&  2193&  2558&  3037&  3390&  3320&  3037&  3110&  2942&  2424&  2586&  2653& 3254& 3838\\
+0.8,+0.9& 905.0&  1352&  1754&  2009&  2222&  2334&  2711&  2870&  2454&  1880&  1391&  1036& 758.7& 544.3& 505.5& 359.6& --- & --- \\
+0.7,+0.8&  1134&  1557&  1781&  1845&  1769&  1823&  1873&  1464& 963.8& 601.6& 339.6& 184.1& 170.1& 230.6&  --- &  --- & --- & --- \\
+0.6,+0.7&  1435&  1455&  1581&  1648&  1791&  1513&  1068& 598.2& 267.2& 155.1& 69.28& 89.01&  --- &  --- &  --- &  --- & --- & --- \\
+0.5,+0.6&  1380&  1372&  1434&  1370&  1201& 870.2& 432.3& 162.2& 71.88& 49.10& 54.01&  --- &  --- &  --- &  --- &  --- & --- & --- \\
+0.4,+0.5&  1477&  1273&  1365&  1369&  1021& 475.5& 161.6& 55.58& 16.32&  --- &  --- &  --- &  --- &  --- &  --- &  --- & --- & --- \\
+0.3,+0.4&  1267&  1154&  1155& 965.3& 574.7& 149.2& 53.26&  --- &  --- &  --- &  --- &  --- &  --- &  --- &  --- &  --- & --- & --- \\
+0.2,+0.3&  1293&  1105&  1041& 742.5& 250.6& 77.66& 110.3&  --- &  --- &  --- &  --- &  --- &  --- &  --- &  --- &  --- & --- & --- \\
+0.1,+0.2&  1351&  1246&  1048& 415.1& 114.3& 41.02&  --- &  --- &  --- &  --- &  --- &  --- &  --- &  --- &  --- &  --- & --- & --- \\
 0.0,+0.1&  1090&  1078& 695.5& 238.2& 45.96&  --- &  --- &  --- &  --- &  --- &  --- &  --- &  --- &  --- &  --- &  --- & --- & --- \\
-0.1, 0.0& 980.4& 783.6& 515.7& 114.6& 20.92&  --- &  --- &  --- &  --- &  --- &  --- &  --- &  --- &  --- &  --- &  --- & --- & --- \\
-0.2,-0.1& 917.7& 746.9& 337.5& 50.92& 3.422&  --- &  --- &  --- &  --- &  --- &  --- &  --- &  --- &  --- &  --- &  --- & --- & --- \\
-0.3,-0.2& 922.7& 586.4& 215.6& 55.88&  --- &  --- &  --- &  --- &  --- &  --- &  --- &  --- &  --- &  --- &  --- &  --- & --- & --- \\
-0.4,-0.3& 698.0& 553.3& 135.3&  --- &  --- &  --- &  --- &  --- &  --- &  --- &  --- &  --- &  --- &  --- &  --- &  --- & --- & --- \\
-0.5,-0.4& 596.9& 482.6& 57.73&  --- &  --- &  --- &  --- &  --- &  --- &  --- &  --- &  --- &  --- &  --- &  --- &  --- & --- & --- \\
-0.6,-0.5& 520.8& 360.7& 34.63&  --- &  --- &  --- &  --- &  --- &  --- &  --- &  --- &  --- &  --- &  --- &  --- &  --- & --- & --- \\
-0.7,-0.6& 450.2& 236.6& 31.22&  --- &  --- &  --- &  --- &  --- &  --- &  --- &  --- &  --- &  --- &  --- &  --- &  --- & --- & --- \\
-0.8,-0.7& 408.8& 184.4&  --- &  --- &  --- &  --- &  --- &  --- &  --- &  --- &  --- &  --- &  --- &  --- &  --- &  --- & --- & --- \\
-0.9,-0.8& 339.7& 107.6&  --- &  --- &  --- &  --- &  --- &  --- &  --- &  --- &  --- &  --- &  --- &  --- &  --- &  --- & --- & --- \\
-1.0,-0.9& 349.8& 63.32&  --- &  --- &  --- &  --- &  --- &  --- &  --- &  --- &  --- &  --- &  --- &  --- &  --- &  --- & --- & --- \\
\hline
\hline
\end{tabular}
\end{center}
\caption{
Shape uncertainty on the MiniBooNE $\nu_\mu$ CCQE flux-integrated double 
differential cross section in units of $10^{-42}~\ucmt/\uGeV$ corresponding 
to Table~\ref{tab:ddsigma}. The total normalization error is $\ttXSerr$\%.}
\label{tab:ddsigma_err}
\end{table*}

\subsection{CCQE-like backgrounds}\label{app:irreducible}
As explained in Sect.~\ref{sec:xsformula}, the CC1$\pip$ interaction with 
intranuclear pion absorption forms a ``CCQE-like'' background in that the 
final state is indistinguishable from the CCQE signal in MiniBooNE.  These 
events originate from the CC1$\pip$ interaction but contain 1 muon and no 
pions in the final state. In the main analysis, this background is subtracted 
to obtain the CCQE observables. In order to facilitate comparisons with models
(or other analyses) that consider all CCQE-like events as CCQE signal, the 
effective double differential cross section for the CC1$\pip$ interaction 
with intranuclear pion absorption is presented in Table~\ref{tab:ddsigma_bkd}.
These values are determined from the \nuance-event generator corrected
to reproduce the MiniBooNE 3-subevent sample and are calculated using 
Eq.~\ref{eq:xseq} with $(d_j-b_j)$ replaced by $b'_j$, the number of 
CCQE-like background events.  A CCQE-like cross section may be 
obtained by adding these numbers (Table~\ref{tab:ddsigma_bkd}) with those 
from Table~\ref{tab:ddsigma}.

\begin{table*}
\begin{center}
\scriptsize\addtolength{\tabcolsep}{-0.8pt}
\begin{tabular}{c|cccccccccccccccccc}
\hline
\hline
\raisebox{-0.5ex}{$\cos\th_\mu$}\raisebox{0.5ex}{$T_\mu$(GeV)}&0.2,0.3&0.3,0.4&0.4,0.5&0.5,0.6&0.6,0.7&0.7,0.8&0.8,0.9&0.9,1.0&
                                              1.0,1.1&1.1,1.2&1.2,1.3&1.3,1.4&1.4,1.5&1.5,1.6&1.6,1.7&1.7,1.8&1.8,1.9&1.9,2.0\\
\hline
+0.9,+1.0&  83.6& 199.8& 285.3& 364.2& 391.1& 403.7& 384.3& 349.2& 301.4& 232.7& 179.2& 136.1& 102.0& 90.73& 76.55& 52.36& 41.47& 54.50\\
+0.8,+0.9& 111.6& 257.4& 351.0& 364.3& 353.2& 288.9& 233.8& 169.5& 106.6& 59.81& 31.21& 20.89& 10.10& 6.008& 2.376& 2.859&  --- &  --- \\
+0.7,+0.8& 118.4& 270.4& 312.6& 280.3& 211.7& 135.7& 81.47& 40.97& 21.56& 9.247& 3.284& 0.875& 0.057&  --- &  --- &  --- &  --- &  --- \\
+0.6,+0.7& 118.9& 260.0& 252.8& 183.4& 101.8& 52.52& 19.75& 7.978& 2.716& 0.281&  --- &  --- &  --- &  --- &  --- &  --- &  --- &  --- \\
+0.5,+0.6& 109.0& 215.2& 181.4& 104.6& 41.87& 16.33& 3.643& 0.492& 0.004&  --- &  --- &  --- &  --- &  --- &  --- &  --- &  --- &  --- \\
+0.4,+0.5& 109.2& 182.0& 122.4& 51.26& 19.76& 4.193& 0.183&  --- &  --- &  --- &  --- &  --- &  --- &  --- &  --- &  --- &  --- &  --- \\
+0.3,+0.4& 104.0& 140.2& 73.71& 24.54& 4.613& 0.151& 0.002&  --- &  --- &  --- &  --- &  --- &  --- &  --- &  --- &  --- &  --- &  --- \\
+0.2,+0.3& 93.84& 107.6& 48.56& 10.78& 0.812&  --- &  --- &  --- &  --- &  --- &  --- &  --- &  --- &  --- &  --- &  --- &  --- &  --- \\
+0.1,+0.2& 76.55& 80.94& 29.02& 3.049& 0.030&  --- &  --- &  --- &  --- &  --- &  --- &  --- &  --- &  --- &  --- &  --- &  --- &  --- \\
 0.0,+0.1& 67.81& 52.89& 13.71& 0.392&  --- &  --- &  --- &  --- &  --- &  --- &  --- &  --- &  --- &  --- &  --- &  --- &  --- &  --- \\
-0.1, 0.0& 58.91& 37.46& 5.565& 0.011&  --- &  --- &  --- &  --- &  --- &  --- &  --- &  --- &  --- &  --- &  --- &  --- &  --- &  --- \\
-0.2,-0.1& 50.47& 22.49& 1.048&  --- &  --- &  --- &  --- &  --- &  --- &  --- &  --- &  --- &  --- &  --- &  --- &  --- &  --- &  --- \\
-0.3,-0.2& 39.03& 12.58& 0.118&  --- &  --- &  --- &  --- &  --- &  --- &  --- &  --- &  --- &  --- &  --- &  --- &  --- &  --- &  --- \\
-0.4,-0.3& 32.41& 7.575& 0.061&  --- &  --- &  --- &  --- &  --- &  --- &  --- &  --- &  --- &  --- &  --- &  --- &  --- &  --- &  --- \\
-0.5,-0.4& 25.72& 2.529& 0.080&  --- &  --- &  --- &  --- &  --- &  --- &  --- &  --- &  --- &  --- &  --- &  --- &  --- &  --- &  --- \\
-0.6,-0.5& 16.78& 1.063& 0.009&  --- &  --- &  --- &  --- &  --- &  --- &  --- &  --- &  --- &  --- &  --- &  --- &  --- &  --- &  --- \\
-0.7,-0.6& 9.963& 0.280& 0.002&  --- &  --- &  --- &  --- &  --- &  --- &  --- &  --- &  --- &  --- &  --- &  --- &  --- &  --- &  --- \\
-0.8,-0.7& 5.005& 0.244&  --- &  --- &  --- &  --- &  --- &  --- &  --- &  --- &  --- &  --- &  --- &  --- &  --- &  --- &  --- &  --- \\
-0.9,-0.8& 4.877& 0.067&  --- &  --- &  --- &  --- &  --- &  --- &  --- &  --- &  --- &  --- &  --- &  --- &  --- &  --- &  --- &  --- \\
-1.0,-0.9& 3.092& 0.013&  --- &  --- &  --- &  --- &  --- &  --- &  --- &  --- &  --- &  --- &  --- &  --- &  --- &  --- &  --- &  --- \\
\hline
\hline
\end{tabular}
\end{center}
\caption{
The predicted $\nu_\mu$ CCQE-like background flux-integrated double differential
cross section in units of $10^{-41}~\ucmt/\uGeV$ corresponding to 
Table~\ref{tab:ddsigma}.}
\label{tab:ddsigma_bkd}
\end{table*}

\subsection{CCQE flux-integrated single differential cross section}\label{app:dsigma}
Table~\ref{tab:dsigma} contains the flux-integrated CCQE single differential 
cross section ($\frac{d\si}{dQ^2_{QE}}$) in bins of $Q^2_{QE}$. $Q^2_{QE}$
is as defined in Eq.~\ref{eq:recQsq}. The shape error and CCQE-like 
background prediction is also reported. The corresponding plot is shown 
in Figure~\ref{fig:dsigma}.  
\begin{table*}
\begin{center}
\begin{tabular}{cccccccc}
\hline
\hline
$Q^2_{QE}$~($\uGeVt$)&&$\frac{d\si}{dQ^2_{QE}}$~($\ucmt/\uGeVt$)&shape error~($\ucmt/\uGeVt$)&CCQE-like bkgd~($\ucmt/\uGeVt$)\\
\hline
0.00-0.05&&$7.681\times 10^{-39}$&$1.493\times 10^{-39}$&$3.876\times 10^{-39}$\\
0.05-0.10&&$1.457\times 10^{-38}$&$1.180\times 10^{-39}$&$3.961\times 10^{-39}$\\
0.10-0.15&&$1.684\times 10^{-38}$&$9.720\times 10^{-40}$&$3.671\times 10^{-39}$\\
0.15-0.20&&$1.703\times 10^{-38}$&$8.216\times 10^{-40}$&$3.064\times 10^{-39}$\\
0.20-0.25&&$1.589\times 10^{-38}$&$5.134\times 10^{-40}$&$2.522\times 10^{-39}$\\
0.25-0.30&&$1.449\times 10^{-38}$&$3.983\times 10^{-40}$&$2.040\times 10^{-39}$\\
0.30-0.35&&$1.329\times 10^{-38}$&$3.386\times 10^{-40}$&$1.633\times 10^{-39}$\\
0.35-0.40&&$1.172\times 10^{-38}$&$2.629\times 10^{-40}$&$1.290\times 10^{-39}$\\
0.40-0.45&&$1.030\times 10^{-38}$&$2.457\times 10^{-40}$&$1.018\times 10^{-39}$\\
0.45-0.50&&$8.852\times 10^{-39}$&$2.975\times 10^{-40}$&$7.874\times 10^{-40}$\\
0.50-0.60&&$7.164\times 10^{-39}$&$3.193\times 10^{-40}$&$5.524\times 10^{-40}$\\
0.60-0.70&&$5.425\times 10^{-39}$&$3.212\times 10^{-40}$&$3.532\times 10^{-40}$\\
0.70-0.80&&$4.032\times 10^{-39}$&$3.442\times 10^{-40}$&$2.302\times 10^{-40}$\\
0.80-1.00&&$2.713\times 10^{-39}$&$2.885\times 10^{-40}$&$1.339\times 10^{-40}$\\
1.00-1.20&&$1.620\times 10^{-39}$&$2.250\times 10^{-40}$&$6.398\times 10^{-41}$\\
1.20-1.50&&$9.915\times 10^{-40}$&$1.407\times 10^{-40}$&$2.466\times 10^{-41}$\\
1.50-2.00&&$5.474\times 10^{-40}$&$2.504\times 10^{-41}$&$3.645\times 10^{-42}$\\
\hline
\hline
\end{tabular}
\end{center}
\caption{
The MiniBooNE $\nu_\mu$ CCQE flux-integrated single differential cross section, 
errors, and predicted CCQE-like background in bins of $Q^2_{QE}$. The total 
normalization error is $\ttXSerr$\%.}
\label{tab:dsigma}
\end{table*}

\subsection{Flux unfolded CCQE cross section}\label{app:sigma}
Table~\ref{tab:sigma} contains the flux-unfolded $\nu_\mu$ CCQE cross 
section values $\si[E_\nu^{QE,RFG}]$ in bins of $E_\nu^{QE,RFG}$.
$E_\nu^{QE,RFG}$ is as defined in Eq.~\ref{eq:recEnu}. The shape and total 
errors along with the CCQE-like background are also reported. The 
corresponding plot is shown in Figure~\ref{fig:sigma}. 

\begin{table*}
\begin{center}
\begin{tabular}{ccccccc}
\hline
\hline
$E_\nu^{QE,RFG}$~(GeV)&&$\si$~($\ucmt$)&shape error~($\ucmt$)&total error~($\ucmt$)&CCQE-like bkgd~($\ucmt$)\\
\hline
0.40-0.45&&$7.985\times 10^{-39}$&$1.997\times 10^{-39}$&$1.997\times 10^{-39}$&$1.731\times 10^{-39}$\\
0.45-0.50&&$8.261\times 10^{-39}$&$1.455\times 10^{-39}$&$1.532\times 10^{-39}$&$1.865\times 10^{-39}$\\
0.50-0.55&&$8.809\times 10^{-39}$&$1.169\times 10^{-39}$&$1.330\times 10^{-39}$&$1.951\times 10^{-39}$\\
0.55-0.60&&$9.530\times 10^{-39}$&$9.537\times 10^{-40}$&$1.209\times 10^{-39}$&$1.978\times 10^{-39}$\\
0.60-0.65&&$1.013\times 10^{-38}$&$7.575\times 10^{-40}$&$1.124\times 10^{-39}$&$1.941\times 10^{-39}$\\
0.65-0.70&&$1.071\times 10^{-38}$&$6.000\times 10^{-40}$&$1.089\times 10^{-39}$&$1.878\times 10^{-39}$\\
0.70-0.75&&$1.111\times 10^{-38}$&$4.496\times 10^{-40}$&$1.065\times 10^{-39}$&$1.770\times 10^{-39}$\\
0.75-0.80&&$1.155\times 10^{-38}$&$3.151\times 10^{-40}$&$1.078\times 10^{-39}$&$1.672\times 10^{-39}$\\
0.80-0.90&&$1.202\times 10^{-38}$&$1.954\times 10^{-40}$&$1.129\times 10^{-39}$&$1.528\times 10^{-39}$\\
0.90-1.00&&$1.230\times 10^{-38}$&$2.714\times 10^{-40}$&$1.217\times 10^{-39}$&$1.334\times 10^{-39}$\\
1.00-1.10&&$1.258\times 10^{-38}$&$4.952\times 10^{-40}$&$1.359\times 10^{-39}$&$1.187\times 10^{-39}$\\
1.10-1.30&&$1.258\times 10^{-38}$&$9.122\times 10^{-40}$&$1.662\times 10^{-39}$&$1.005\times 10^{-39}$\\
1.30-1.50&&$1.278\times 10^{-38}$&$1.417\times 10^{-39}$&$2.116\times 10^{-39}$&$7.944\times 10^{-40}$\\
1.50-2.00&&$1.236\times 10^{-38}$&$1.991\times 10^{-39}$&$2.613\times 10^{-39}$&$4.822\times 10^{-40}$\\
\hline
\hline
\end{tabular}
\end{center}
\caption{
The MiniBooNE $\nu_\mu$ CCQE flux-unfolded cross section, errors, and 
predicted CCQE-like background in bins of $E_\nu^{QE,RFG}$.}
\label{tab:sigma}
\end{table*}

\end{document}